\documentclass[12pt]{article}
\usepackage[colorlinks=true, urlcolor=navyblue, linkcolor=navyblue, citecolor=navyblue]{hyperref}
\usepackage{graphicx,amsmath,amssymb}
\usepackage{indentfirst}
\usepackage[usenames]{color}
\usepackage{epstopdf}
\usepackage{enumerate}
\usepackage{appendix}
\usepackage{lineno}
\usepackage{setspace}
\usepackage{ulem}

\definecolor{navyblue}{rgb}{0,0.08,0.45}
\definecolor{darkred}{rgb}{0.7,0.0,0.0}
\definecolor{darkgreen}{rgb}{0,0.6,0.2}

\newcommand{\beq}{\begin{equation}}
\newcommand{\enq}{\end{equation}}
\newcommand{\beqa}{\begin{eqnarray}}
\newcommand{\beqast}{\begin{eqnarray*}}
\newcommand{\enqa}{\end{eqnarray}}
\newcommand{\enqast}{\end{eqnarray*}}
\newcommand{\nn}{\nonumber}

\newcommand{\bec}{\begin{center}}
\newcommand{\enc}{\end{center}}
\newcommand{\beqo}{\begin{quote}}
\newcommand{\enqo}{\end{quote}}
\newcommand{\bem}{\begin{minipage}}
\newcommand{\enm}{\end{minipage}}

\newcommand{\mbf}{\mathbf}

\newcommand{\req}[1]{(\ref{#1})}

\newcommand{\half}{\textstyle \frac{1}{2}}

\newcommand{\cL}{{\cal L}}

\newcommand{\de}{\delta}

\newcommand{\la}{\lambda}
\newcommand{\rh}{\rho}
\newcommand{\si}{\sigma}

\newcommand{\ph}{\phi}

\newcommand{\De}{\Delta}

\newcommand{\La}{\Lambda}

\newcommand{\Si}{\Sigma}

\newcommand{\ud}{\mathrm{d}}

\definecolor{green}{rgb}{0,.5,0}

\begin{document}

\begin{flushright}
{\small SLAC-PUB-16545
}
\end{flushright}

\vspace{50pt}

\begin{center}

{\huge  Meson/Baryon/Tetraquark Supersymmetry\\}

\vspace{3pt}

{\huge   from  Superconformal Algebra and Light-Front Holography}
\end{center}

\vspace{30pt}

\centerline{Stanley J. Brodsky}

\vspace{3pt}

\centerline {\it SLAC National Accelerator Laboratory, Stanford
University, Stanford, CA 94309,
USA
}

\vspace{10pt}

\centerline{Guy F. de T\'eramond}

\vspace{3pt}

\centerline {\it Universidad de Costa Rica, 11501 San Pedro de
Montes de Oca, Costa
Rica
}

\vspace{10pt}

\centerline{Hans G\"unter Dosch}

\vspace{3pt}

\centerline{\it Institut f\"ur Theoretische Physik der Universit\"at,}

\centerline{\it   Philosophenweg
16, D-69120 Heidelberg,
Germany
}

\vspace{10pt}

\centerline{C\'edric Lorc\'e}

\vspace{3pt}

\centerline{\it {Centre de Physique Th\'eorique, \'Ecole
polytechnique, CNRS, Universit\'e Paris-Saclay,}}

\centerline{\it {F-91128 Palaiseau, France}
}

\vspace{15pt}

{\small
}

\vspace{60pt}

{\small
\centerline{\href{mailto:sjbth@slac.stanford.edu}{\tt sjbth@slac.stanford.edu}}
}

{\small
\centerline{\href{mailto:gdt@asterix.crnet.cr}{\tt gdt@asterix.crnet.cr}}
}

{\small
\centerline{\href{mailto:dosch@thphys.uni-heidelberg.de}{\tt dosch@thphys.uni-heidelberg.de}}
}

{\small
\centerline{\href{mailto:cedric.lorce@polytechnique.edu}{\tt cedric.lorce@polytechnique.edu}}
}


\newpage

\vspace{15pt}

\begin{abstract}

\vspace{15pt}

Superconformal algebra leads to  remarkable connections between the masses of mesons and baryons of the same parity -- supersymmetric relations between the bosonic  and fermionic bound states of QCD.   Supercharges connect the mesonic eigenstates  to their baryonic superpartners, where the mesons have internal  angular momentum one unit higher than the baryons:  $L_M = L_B+1.$
The dynamics of  the superpartner hadrons also match; for example, the power-law fall-off of the form factors are the same for the mesonic and baryonic superpartners, in agreement with twist counting rules.
An effective supersymmetric  light-front Hamiltonian for hadrons composed of light quarks can be constructed by embedding superconformal quantum mechanics into AdS space.   This procedure also generates  a spin-spin interaction between the hadronic constituents.  A specific breaking of conformal symmetry inside the graded algebra determines a unique quark-confining  light-front potential for light hadrons in agreement with  the soft-wall AdS/QCD approach and light-front holography.
Only one mass parameter $ \sqrt \lambda$ appears; it sets the confinement  mass scale, a universal value for the slope of all Regge  trajectories,  the  nonzero mass of the proton and other hadrons  in the  chiral limit, as well as the length scale which underlies their structure.  The mass  for the pion eigenstate  vanishes in the chiral limit. 
When one includes the  constituent quark masses using the Feynman-Hellman theorem, the predictions are consistent with the empirical features of the light-quark hadronic spectra.     Our analysis can be consistently applied to the excitation spectra of the $\pi, \rho, K, K^*$ and $\phi$   meson families as well as to the $N, \Delta, \Lambda, \Sigma, \Sigma^*, \Xi$ and $\Xi^*$ baryons.  We also predict the existence of tetraquarks which are degenerate in mass with baryons with the same angular momentum.   The mass-squared  of the light hadrons can be expressed in a universal and frame-independent decomposition of contributions from the constituent kinetic energy, the confinement potential,  and  spin-spin contributions.    We also predict features of hadron dynamics, including  hadronic light-front wavefunctions, distribution amplitudes, form factors, valence structure functions and  vector meson electroproduction phenomenology.  The mass scale $ \sqrt \lambda$ can be connected to the parameter   $\Lambda_{\overline {MS}}$ in the QCD running coupling by matching the nonperturbative dynamics, as described by the light-front holographic approach.
to the perturbative QCD  regime. The result is an effective coupling  defined at all momenta.       The 
matching of the high and low momentum-transfer regimes determines a scale $Q_0$ proportional to $ \sqrt \lambda$   which  sets the interface between perturbative and nonperturbative hadron dynamics.  
The use of $Q_0$ to  resolve  the factorization scale uncertainty for structure functions and distribution amplitudes,  in combination with the scheme-independent Principle of Maximal Conformality (PMC)  procedure for  setting  renormalization scales,  can 
greatly improve the precision of perturbative QCD predictions.

\end{abstract}

\newpage

\section{Introduction}

\begin{figure}
 \begin{center}
\includegraphics[height= 12cm,width=15cm]{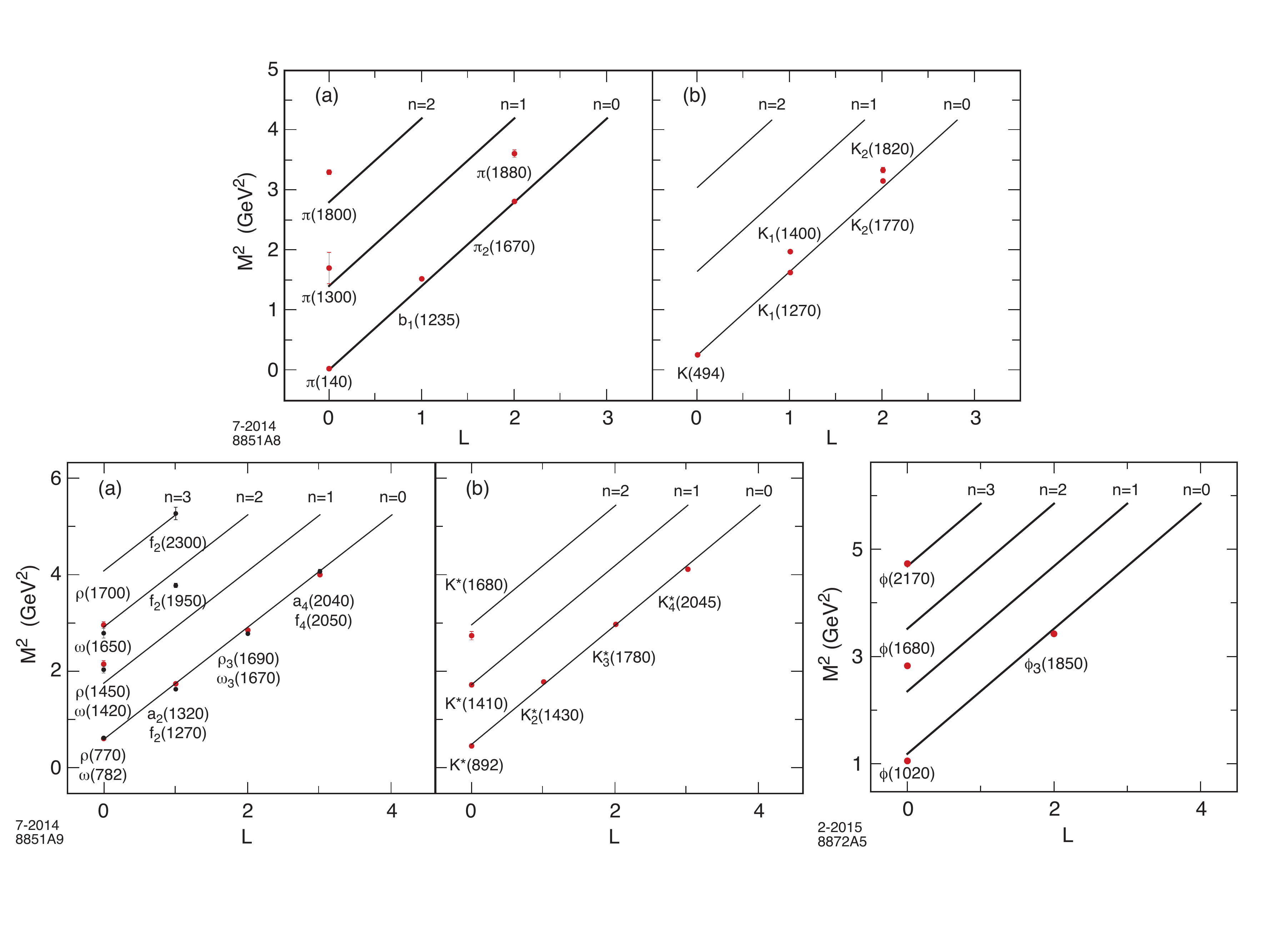}
\end{center}
\caption{Comparison of the 
light-front holographic prediction~\cite{deTeramond:2008ht}  $M^2(n, L, S) = 4\lambda(n+L +S/2)$ for the orbital $L$ and radial $n$ excitations of the meson spectrum with experiment. See Ref.~\cite{Brodsky:2014yha}}
\label{ReggePlot} 
\end{figure}

\begin{figure}
 \begin{center}
 \includegraphics[height=6cm,width=7.5cm]{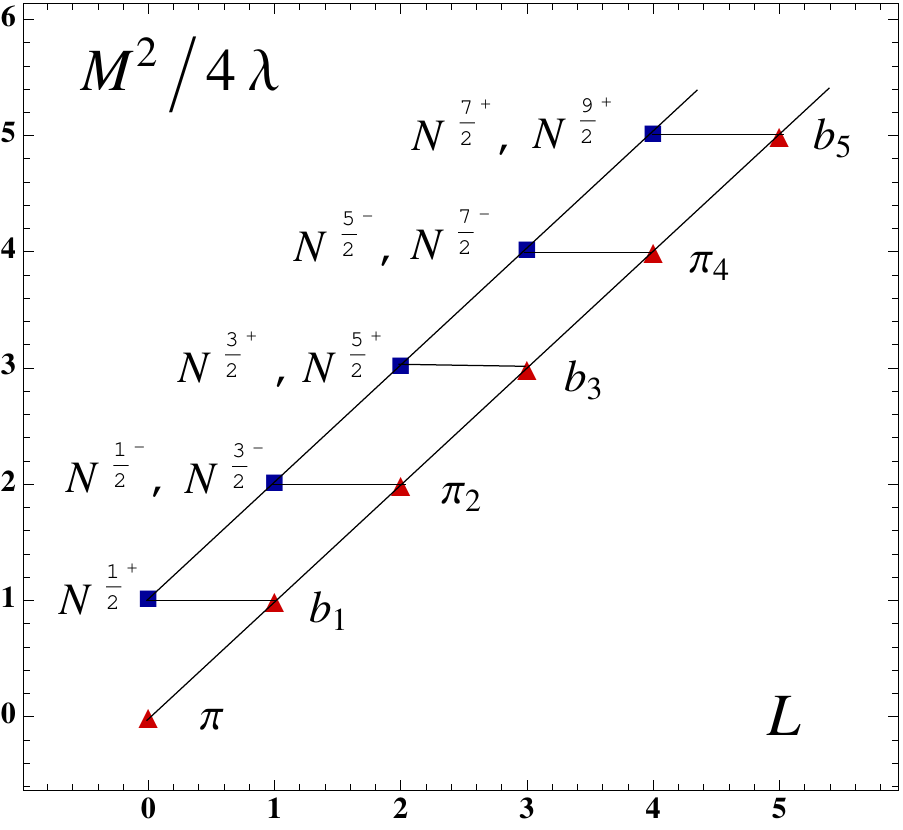}
 \includegraphics[height=6cm,width= 7.5cm]{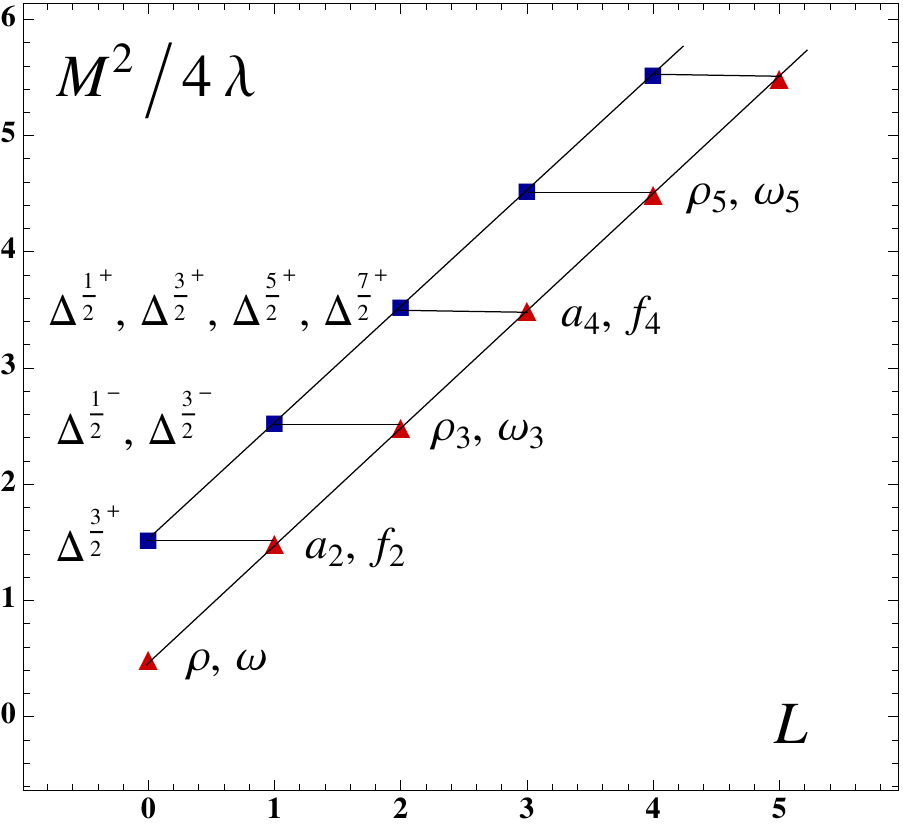}
\end{center}
\caption{Comparison of Regge trajectories for meson and baryons. The Regge slopes in $L$ are 
 predicted to be equal for both the meson and baryon trajectories. See Refs.~\cite{deTeramond:2014asa,Dosch:2015nwa}}
\label{Reggeslopes}
\end{figure}

\begin{figure}[ht]
\setlength{\unitlength}{1mm}
\begin{center}
\includegraphics[width=15cm] {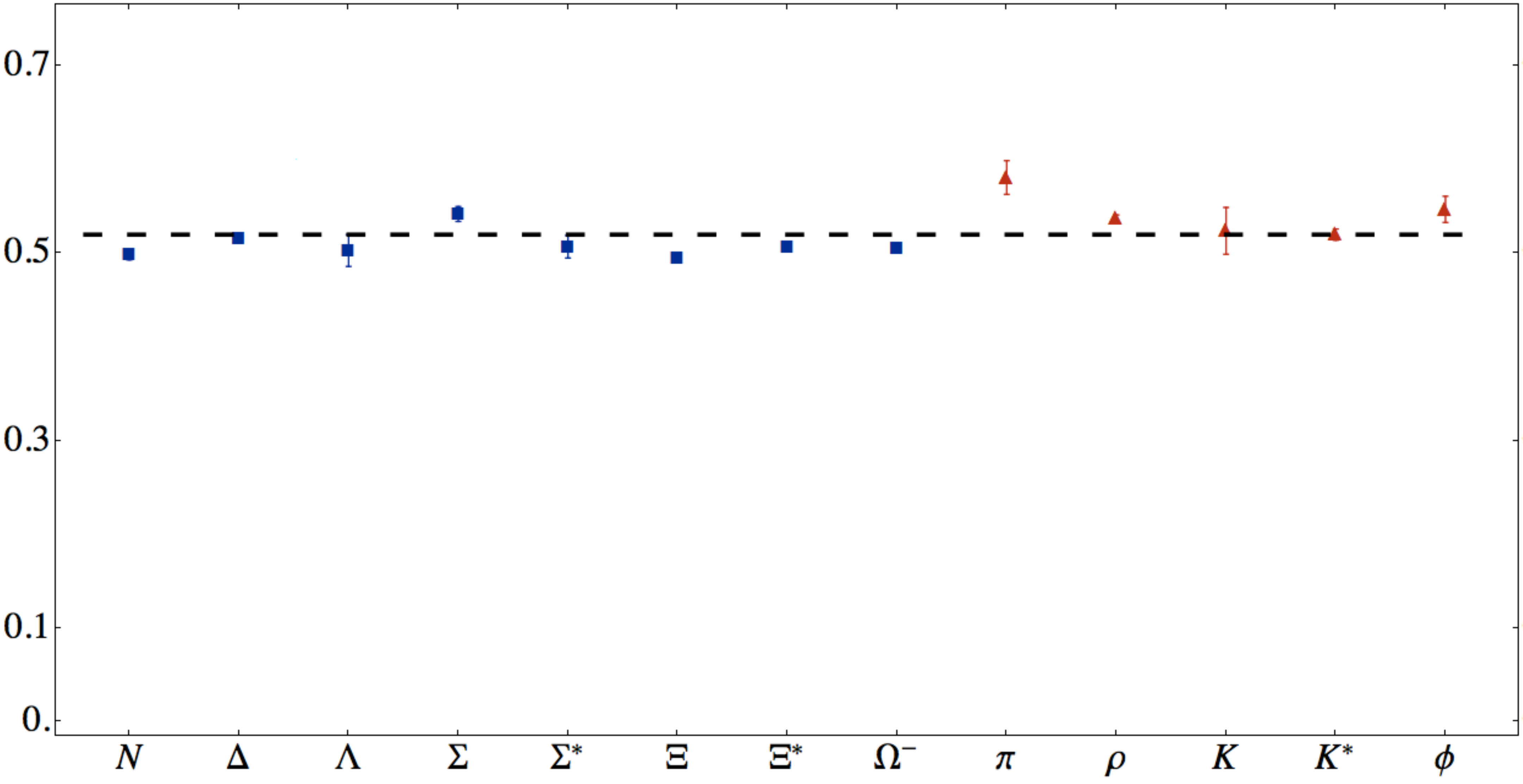}
\end{center}
\caption{\label{slope} Best fit for the value of the slope of the different Regge trajectories for baryons and mesons including all radial and orbital excitations. The dotted line is the average value $\kappa = \sqrt{\la}= 0.523$ GeV; it has the standard deviation  $\si =0.024$ GeV. For the  baryon sample alone the values are $0.509 \pm  0.015$ GeV and for the mesons $0.524 \pm 0.023$ GeV.  See. Ref.~\cite{Brodsky:2016yod}}
\label{Reggeslopefit}
\end{figure}

A remarkable empirical feature of the hadronic spectrum is the near equality of the slopes of meson and baryon Regge trajectories.  The  square of the masses of hadrons composed of light quarks is linearly proportional not only to $L$, the orbital angular momentum, but also to the principal quantum number $n$, the number of radial nodes in the hadronic wavefunction as seen in  Fig.~\ref{ReggePlot}.    The Regge slopes in $n$ and $L$ are equal, as in the meson formula  
$M^2_M(n,L, S) = 4 \lambda(n+L + S/2$ from 
light front holographic QCD~\cite{deTeramond:2008ht},  
but even more surprising, they are observed to be equal for both the meson and baryon trajectories, 
as shown in Fig.~\ref {Reggeslopes}.  The mean value for all of the slopes is $\kappa = \sqrt{\la}= 0.523$ GeV.  See Fig.~\ref {Reggeslopefit}.

These striking features of hadron spectroscopy are difficult to understand if one assumes that mesons are $q \bar q$ bound states and baryons are composites of three quark constituents.  However, there is a simple physical explanation why the baryon and meson spectra could be similar in QCD.  Suppose that  the two color-triplet $3_C$ quarks in a baryon bind to form an anti-color-triplet $\bar 3_C$  diquark ``cluster".    This  $[q q]_{\bar 3_C}$   diquark state  could then bind to  the remaining $q_{3_C} $ quark  to  form the baryonic color singlet.   The color-binding diquark-quark interactions of  the $([q q]_{{\bar 3}_C} + q_{3_C})$   baryon  would  then mimic the 
$(\bar q_{\bar 3_C}  + q _{3_C})$ color-binding of a meson.   The relative orbital angular momentum $L$ in the $q \bar q$ meson would have its counterpart in the relative orbital angular momentum $L$ between the quark and diquark cluster in the baryon.  The diquark cluster can have spin $S=0$ or $S=1,$   leading, respectively, to the spin-${1\over 2}$ nucleon and the spin-${3\over 2}$ $\Delta$ states and their radial and orbital excitations.

Note that if this simple picture of hadron structure is correct,  then a  $\bar 3_C$ diquark $qq$ cluster and a $3_C$ antidiquark $\bar q \bar q$  cluster should  also bind to form  $[q  q]_{\bar 3C} [\bar q \bar q]_{3C}$  color-singlet tetraquarks. 

\begin{figure}
 \begin{center}
 \includegraphics[height=11cm,width=15cm]{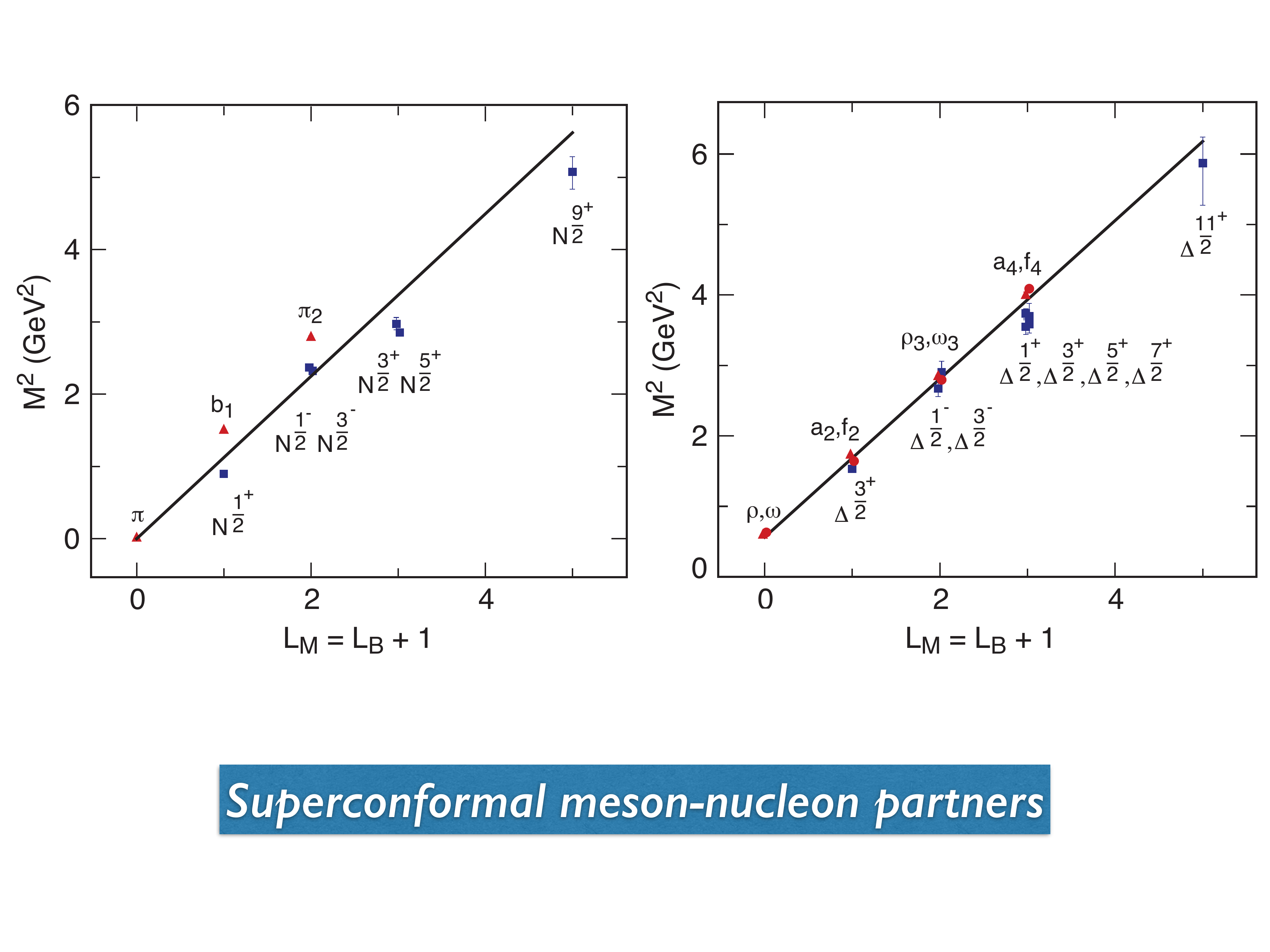}
\end{center}
\caption{Comparisons of the  meson and the  baryon Regge trajectories.   Superconformal algebra  predicts the degeneracy of the  meson and baryon trajectories if one identifies a meson with internal orbital angular momentum $L_M$ with its superpartner baryon with $L_M = L_B+1$.
See Refs.{~\cite{deTeramond:2014asa,Dosch:2015nwa}}}
\label{SUSYduo}
\end{figure}

There is in fact another surprising similarity of the observed meson and baryon spectra.  Suppose we shift the relative angular momentum of mesons $L_M$ versus  the relative angular momentum of baryons $L_B$ by one unit; {\it i.e.}, we will compare the masses of mesons and baryons with $L_M= L_B+1$.    Remarkably the empirical masses of the mesons and baryons match each other very well.   See Fig.~\ref {SUSYduo}.  The equal masses of mesons and baryons with with $L_M= L_B+1$ can thus be considered as a supersymmetric feature of hadron physics, a relation between bosons and fermions. 

The dynamics of mesonic and baryonic  superpartners also match; {\it e.g.},  the counting rules for the fall-off of their form factors are identical. The pion  has $L_M=0$, and thus it has no baryon counterpart. Notice that the shifted mesons and baryons have the same parity and the same twist $\tau_B = 3 + L_B = \tau_M  = 2 + L_M$,  where $\tau$ is the index  in the operator product expansion which controls hadron wavefunctions at short distances $x^2 \to 0$;     thus the power-law fall-off of  form factors and other exclusive scattering amplitudes at high momentum transfer of the meson and its baryon partner will be identical.

As we shall discuss in this contribution, these empirical observations
are in fact predictions of {\it superconformal algebra}  and {\it light-front holography}.  See Ref.~\cite{Brodsky:2016yod}.   Each hadron is an eigenstate of a bound-state light-front Schr\"odinger  equation  with a unique color-confining potential.   The mass-squared  of light hadrons can be expressed in a universal and frame-independent decomposition. The light-front (LF) formalism is Lorentz invariant -- independent of the Lorentz frame of the observer.  Moreover, the color-confining potential that appears in the  LF Hamiltonian which produces the observed spectrum has a unique analytic form.  

Light-Front  Quantization~\cite{Dirac:1949cp} --  Dirac's ``Front Form" -- provides a physical, frame-independent formalism for hadron dynamics and structure.  Observations  of an object such as a flash photograph or deep inelastic lepton-proton scattering are made along the front of a light wave; {\it i.e.}, at fixed LF time $x^+ = x^0 + x^3/c$. Observables such as structure functions, transverse momentum distributions, and distribution amplitudes are defined from the hadronic light-front wavefunctions.   One can obtain new insights into the hadronic 
spectrum, light-front wavefunctions, and the functional form of the QCD running coupling in the nonperturbative domain using light-front holography -- the duality between the front form and AdS$_5$, the space of isometries of the conformal group.  The LF formalism for QCD is reviewed in Ref.~\cite{Brodsky:1997de}. 
For an introduction, see Ref.~\cite{Brodsky:1981jv}.

The  semiclassical LF effective theory based on superconformal quantum mechanics and light-front holography also captures other essential features of hadron physics expected from confined quarks in QCD and its chiral properties. A mass scale emerges from a nominal conformal theory, and  a massless  pseudoscalar $q \bar q$ bound state -- the  pion -- appears in the limit of zero-quark masses. Moreover, the eigenvalues of the resulting light-front Hamiltonian predict the same slope for Regge trajectories in both $n$,  the radial excitations,  and $L$, the orbital excitations, as  observed empirically.  This nontrivial aspect of hadron physics~\cite{Glozman:2007at, Shifman:2007xn}  -- the observed equal slopes of the radial and angular Regge trajectories -- is  also a property of the Veneziano dual amplitude~\cite{Veneziano:1968yb}.

The one-dimensional representation of the superconformal algebra  in one dimension, superconformal quantum mechanics~\cite{Fubini:1984hf},  has
four mass-degenerate supersymmetric components: a meson with orbital angular momentum $L_M$,    its baryon superpartner with two orbital angular momentum components  $L_B= L_M-1$ and $ L_B+1$, with equal weight and mass;  these two components correspond to the quark spin $S^z= \pm 1/2$, parallel or anti-parallel to the baryon spin $J^z$;  and the fourth component  --  a bosonic tetraquark diquark-antidiquark partner with $L_T=L_B$~\cite{Brodsky:2016yod}.  The physical picture that baryons are effectively bound states of a quark and diquark cluster  and tetraquarks are diquark-antidiquark bound states underlie this approach.    In fact, as we shall show, the resulting tetraquarks will be degenerate in mass with baryons with the same angular momentum. The existence of spin $J=0,1$ tetraquarks, plus their orbital and  radial excitations, is also necessary consequence of the superconformal algebra approach.

The equal weight and mass  of the $L_B$ and $L_B+1$ components of the baryon corresponds to a feature of chiral symmetry in the quark-diquark LF wavefunction.   The spin of the proton is 
carried by the quark's orbital angular momentum: $J^z= \langle L^z\rangle= {1\over 2}[L^z=1 +   L^z=0] = 1/2$, not the quark spin.  This result is also a feature of the Skyrme model~\cite{Brodsky:1988ip}.

It is also remarkable that the superconformal algebra formalism which underlies the supersymmetric and chiral properties of hadron physics is also consistent with the QCD
light-front holographic approach to hadron physics with a specific soft-wall dilaton.   The same approach, using light-front holography, dictates the behavior of the QCD running coupling and its $\beta$ function in the infrared nonperturbative regime.  The mass scale underlying confinement and hadron masses  can be connected to the parameter   $\Lambda_{\overline {MS}}$ in the QCD running coupling by matching the nonperturbative dynamics, as described by  
holographic QCD, to the perturbative QCD  regime. The result is an effective coupling  defined at all momenta.      The 
matching of the high and low momentum transfer regimes also determines a mass scale $Q_0$ which  sets the interface between perturbative and nonperturbative hadron dynamics.  
The use of $Q_0$ to  resolve  the factorization scale uncertainty for structure functions and distribution amplitudes,  in combination with the principle of maximal conformality (PMC)  for  setting the  renormalization scales,  can 
greatly improve the precision of perturbative QCD predictions for collider phenomenology.

\section{Color Confinement and Supersymmetry in Hadron Physics from LF Holography}

A key problem in hadron physics is how to obtain a color-confining model  for hadron physics which can predict both the hadron spectrum and the hadronic light-front wave functions LFWFs.  
If one neglects the Higgs couplings of quarks, then no mass parameter appears in the QCD Lagrangian, and the theory is conformal at the classical level.   Nevertheless,  hadrons have a finite mass.  In a remarkable paper,  de Alfaro, Fubini and Furlan  (dAFF)~\cite{deAlfaro:1976je}  showed in a 1+ 1 quantum mechanical model
that a mass gap and a mass scale  $\kappa$ may appear in the Hamiltonian and the equations of motion without affecting the conformal invariance of the action. 
A scale is introduced by constructing a generalized Hamiltonian $G$ which is a superposition of the original Hamiltonian $H$, the dilatation generator $D$, and the generator of special conformal transformations $K$~\cite{deAlfaro:1976je}. The new hamiltonian $G$  preserves the quantum-mechanical evolution and the conformal invariance of the action.

The dAFF  method can be extended~\cite{Brodsky:2013ar}   to relativistic frame-independent light-front Hamiltonian theory.  Remarkably, the resulting light-front potential has a unique form of a harmonic oscillator $\kappa^4 \zeta^2$ in the 
light-front invariant impact variable $\zeta$ where $ \zeta^2Ê = b^2_\perp x(1-x)$.  The mass parameter $\kappa = \sqrt \lambda$ can be considered  as the fundamental mass scale of QCD, but it is not determined  in absolute units, since QCD does have knowledge of conventional units such as $MeV.$  It is a place holder. What is predicted are ratios, such as the ratio of the proton mass to the  $\rho$ mass, the ratio of the Regge slope to the proton mass,  hadron radii times hadron masses, the ratio of  the scheme-dependent $\Lambda_s $, which controls the pQCD running coupling (in the $\overline{MS}$ or any other renormalization  scheme), to the proton mass.  One thus retains dilatation invariance when one rescales  $\kappa \to C \kappa$.  In addition a new time variable appears with finite range reflecting the limited  difference between LF times when one observes the components of a finite size hadron.

The result is  a single-variable frame-independent relativistic equation of motion for  $q \bar q $ mesonic bound states, 
a ``Light-Front Schr\"odinger Equation"~\cite{deTeramond:2008ht}, analogous to the nonrelativistic radial Schr\"odinger equation in quantum mechanics.  The  Light-Front Schr\"odinger Equation  incorporates color confinement and other essential spectroscopic and dynamical features of hadron physics, including a massless pion for zero quark mass and linear Regge trajectories with the same slope  in the radial quantum number $n$   and internal  orbital angular momentum $L$.   

The same light-front  equation for mesons of arbitrary spin $J$ can be derived~\cite{deTeramond:2013it}
from the holographic mapping of  the ``soft-wall model" modification of AdS$_5$ space with the specific dilaton profile $e^{+\kappa^2 z^2},$  where one identifies the fifth dimension coordinate $z$ with the light-front coordinate $\zeta$.  The five-dimensional AdS$_5$ space provides a geometrical representation of the conformal group.
It is holographically dual to 3+1  spacetime at fixed light-front time $\tau = t+ z/c$.  

\begin{figure}
 \begin{center}
\includegraphics[height= 10cm,width=15cm]{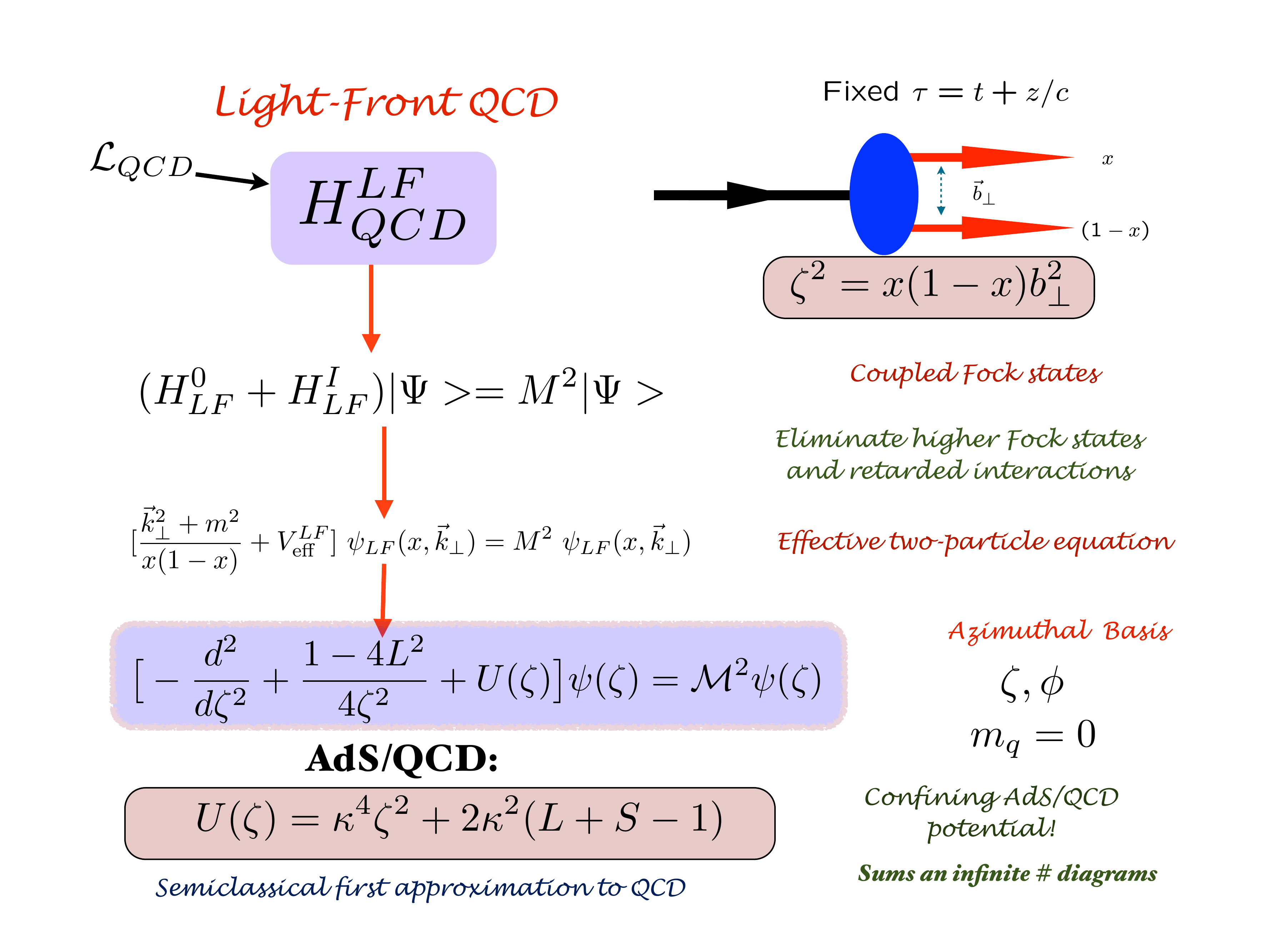}
\end{center}
\caption{Derivation of the Effective Light-Front Schr\"odinger Equation from QCD.  As in QED, one reduces the LF Heisenberg equation $H_{LF}|\Psi >   = M^2 |\Psi>$ 
to an effective two-body eigenvalue equation for $q \bar q$ mesons by systematically eliminating higher Fock states. One utilizes the LF radial variable $\zeta$, where $\zeta^2 = x(1-x)b^2_\perp$ is conjugate to the $q \bar q$ LF kinetic energy $k^2_\perp\over x(1-x)$ for $m_q=0$. This allows the reduction of the dynamics to a single-variable bound state equation acting on the valence $q \bar q$ Fock state.  The confining potential $U(\zeta)$, including its spin-$J$ dependence, is derived from the soft-wall AdS/QCD model with the dilaton  $e^{+\kappa^2 z^2 },$ where $z$ is the fifth coordinate of AdS$_5$ holographically dual  to $\zeta$. See Ref.~\cite{deTeramond:2013it}.   The resulting light-front harmonic oscillator confinement potential $\kappa^4 \zeta^2 $ for light quarks is equivalent to a linear confining potential for heavy quarks in the instant form~\cite{Trawinski:2014msa}. }
\label{FigsJlabProcFig2.pdf}
\end{figure}

The derivation of the confining LF Schrodinger Equation is outlined in Fig. \ref{FigsJlabProcFig2.pdf}. 
The reduction to an effective Hamiltonian acting on the valence Fock state of hadrons in QCD is analogous to the reduction used in precision analyses in QED for atomic physics.

The  combination of light-front dynamics, its holographic mapping to AdS$_5$ space, and the dAFF procedure give new  insight into the physics underlying color confinement, the nonperturbative QCD coupling, and the QCD mass scale.  A comprehensive review is given in  Ref.~\cite{Brodsky:2014yha}.  The $q \bar q$ mesons and their valence LF wavefunctions are the eigensolutions of the frame-independent relativistic bound state LF Schr\"odinger equation.  The mesonic $q\bar  q$ bound-state eigenvalues for massless quarks are $M^2(n, L, S) = 4\kappa^2(n+L +S/2)$.
The equation predicts that the pion eigenstate  $n=L=S=0$ is massless at zero quark mass.    The  Regge spectra of the pseudoscalar $S=0$  and vector $S=1$  mesons  are 
predicted correctly, with equal slope in the principal quantum number $n$ and the internal orbital angular momentum $L$.  
The comparison with experiment is shown in Fig.~\ref{ReggePlot}.

\begin{figure}
\begin{center}
\includegraphics[height=11cm,width=15cm]{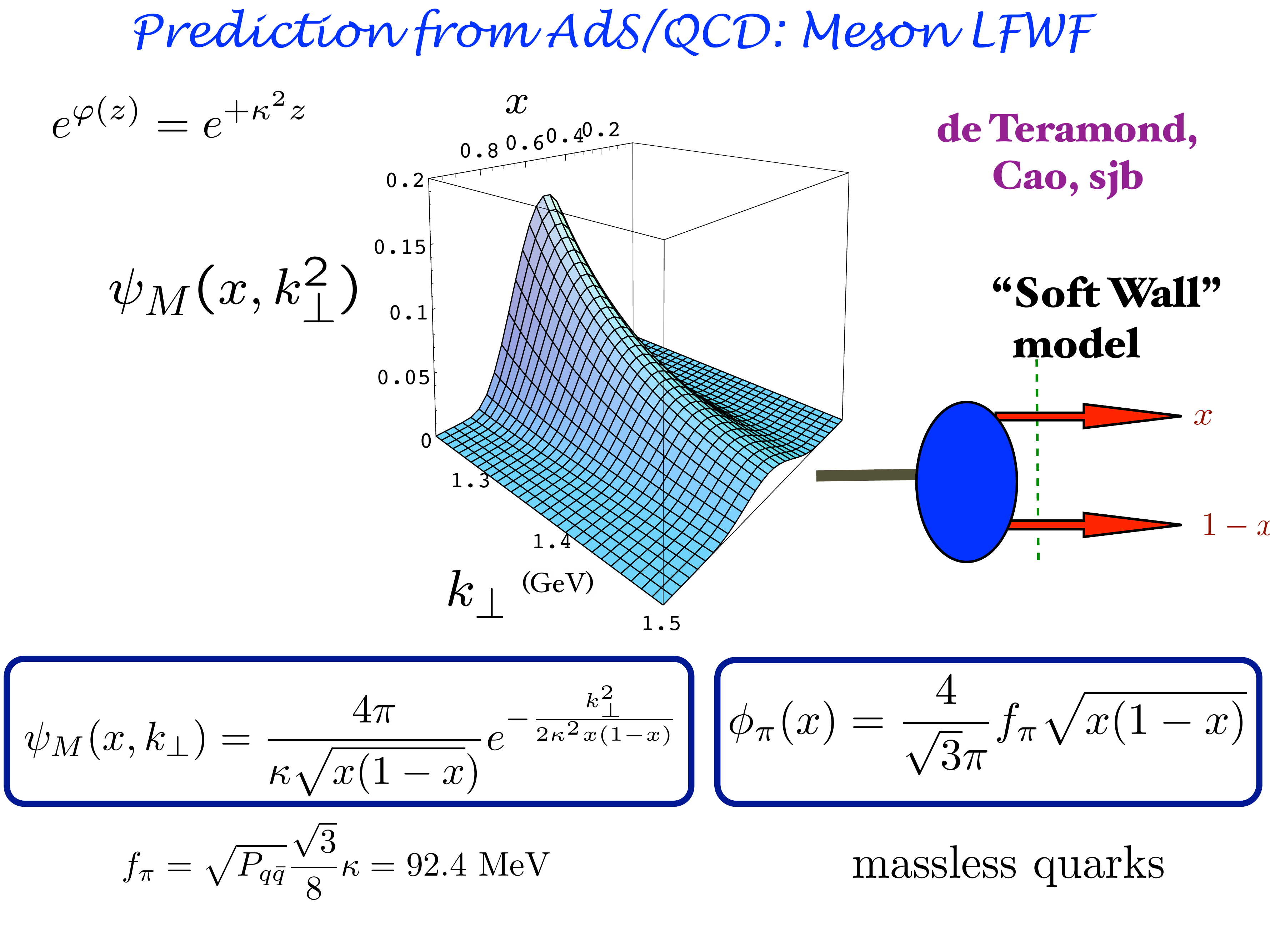}
\end{center}
\caption{Prediction from AdS/QCD and Light-Front Holography for  meson LFWFs  $\psi_M(x,\vec k_\perp)$   and the pion distribution amplitude.     
}
\label{MesonLFWF.pdf}
\end{figure}

The AdS/QCD light-front holographic eigenfunction for the vector meson LFWFs $\psi_V(x,\vec k_\perp)$ gives excellent 
predictions for the observed features of diffractive $\rho$ and $\phi$ electroproduction~\cite{Forshaw:2012im,Ahmady:2016ujw}.  
Note that the prediction for the LFWF is a function of the LF kinetic energy $\vec k^2_\perp/ x(1-x)$ -- the conjugate of the LF radial variable $\zeta^2 = b^2_\perp x(1-x)$ -- times a function of $x(1-x)$. It does not factorize as a  function of $\vec k^2_\perp$ times a function of $x$.  The resulting  nonperturbative pion distribution amplitude $\phi_\pi(x) = \int d^2 \vec k_\perp \psi_\pi(x,\vec k_\perp) = (4/  \sqrt 3 \pi)  f_\pi \sqrt{x(1-x)}$,  see Fig. \ref{MesonLFWF.pdf}, is  consistent with the Belle data for the photon-to-pion transition form factor~\cite{Brodsky:2011xx}.

\subsection{Construction of the Hamiltonian from superconformal algebra~\cite{Brodsky:2016yod}}

The essential features of a simple superconformal graded algebra~\cite{Fubini:1984hf}  in one dimension, {\it conf}({$\mathbb R^1$}) are based on the  generators of translation, dilatation and the special conformal transformation $H$, $D$ and $K$, respectively.   
A mass scale $\kappa = \sqrt \lambda$ appears when one modifies the superconformal operator by an operator $S$ where 
$K = {1\over 2} [S, S^\dagger]$ is the special conformal operator.

By introducing the supercharges $Q$, $Q^\dagger$, $S$ and $S^\dagger$, one constructs the extended algebraic structure~\cite{Fubini:1984hf, Haag:1974qh}  with the relations
\begin{align} \label{susy-extended}  \nn
\half\{Q,Q^\dagger\} &= H, & \half\{S,S^\dagger\}&=K, \\ 
\{Q,S^\dagger\} &= { f} \, { {\mathbf I}} - B + 2 i D, &
\{Q^\dagger,S\} &= { f} \, {{\mathbf I}} - B - 2 i D,
\end{align}
where $f$ is a real number, ${\mathbf I}$ is the identity operator,
$B=\frac{1}{2}[\psi^\dag,\psi]$ is a bosonic operator with
$\{\psi,\psi^\dag\}={\mathbf I}$, $S=x\,\psi^\dag $ and
$S^\dag=x\, \psi $. The operators $H$, $D$ and $K$ satisfy the
conformal algebra 
\beq [H,D]= iH, \qquad [H,K]= 2 i D, \qquad
[K,D]=-i K . 
\enq

The fermionic operators can be conveniently represented in a spinorial space $S_2={\cal L}_2({\mathbb R^1})\otimes {\mathbb C^2}$ as 2$\times$2 matrices 
\beq
Q = \left(\begin{array}{cc}
0&q\\
0&0\\
\end{array}
\right) ,
\quad Q^\dagger=\left(\begin{array}{cc}
0&0\\
q^\dagger&0
\end{array}
\right),
\enq
\beq
S = \left(\begin{array}{cc} 
0& x\\
0&0
\end{array}
\right), \quad  S^\dagger=\left(\begin{array}{cc}
0&0\\
x&0\\
\end{array}
\right) , \enq 
    with 
      \beq \label{q} q =-\frac{\ud}{\ud x} + \frac{f}{x},
\qquad
 q^\dagger = \frac{\ud}{\ud x}  + \frac{f}{x}.
\enq

Following the analysis of Fubini and Rabinovici~\cite{Fubini:1984hf}, which extends the treatment of the conformal group by de Alfaro, Fubini and Furlan~\cite{deAlfaro:1976je} to supersymmetry, we construct ~\cite{deTeramond:2014asa, Dosch:2015nwa} a generalized Hamiltonian from the supercharges
\beq \label{R}
R_\la =Q+\la\,S =\left(\begin{array}{cc}
0&-\frac{\ud}{\ud x} + \frac{f}{x}+\la\,x\\
0&0\\
\end{array}
\right) ,
\enq
\beq  \label{Rdag}
\quad R_\la^ \dagger=Q^\dagger + \la S^\dagger =\left(\begin{array}{cc}
0&0\\
\frac{\ud}{\ud x} + \frac{f}{x}+\la\,x&0
\end{array}
\right),
\enq
namely
\beq \label{G}
G = \{R_\la, R_\la^\dagger\}.
\enq
Since the dimensions of the generators $Q$ and $S$ are different, a scale $\la$, with dimensions of mass squared,  is introduced in the Hamiltonian in analogy with the earlier treatment of conformal quantum mechanics given in Ref.~\cite{deAlfaro:1976je}.   As shown by de Alfaro, Fubini and Furlan~\cite{deAlfaro:1976je}, the conformal symmetry of the action is retained despite the presence of a mass scale in the Hamiltonian.

The supercharges and the new Hamiltonian $G$  satisfy, by construction,  the relations:
\beq  
\label{newrel} \{R_\la^\dagger,R_\la^\dagger\}
=\{R_\la,R_\la\}=0,\qquad[R_\la,G]= [R_\la^\dagger,G]=0,
\enq
which, together with Eq. \req{G}, close a graded algebra $sl(1/1)$, as in Witten's supersymmetric quantum mechanics~\cite{Witten:1981nf}.  Since the Hamiltonian $G$ commutes with $R^\dagger_\la$, it  follows that the states  $\vert \phi \rangle$ and $R^\dagger  \vert \phi \rangle$ have identical eigenvalues. Furthermore, it follows that if $|\ph_0 \rangle$ is an eigenvalue of $G$ with zero eigenvalue, it is annihilated by the operator $R_\la^\dagger$:
\beq  \label{anni} 
R_\la^\dagger|\ph_0 \rangle = 0.
\enq

In matrix representation Eq. \req{G} is given by
\beq
G = 2 H + 2 \la^2 K + 2 \la \left( f\, \mbf{I} - \si_3 \right),
 \quad \mbox{with} \quad \si_3 = \left(\begin{array}{cc} 1&0 \\ 0 & -1 \end{array}\right).
\enq
The new Hamiltonian is diagonal, with elements:
\beqa \label{Gsc}
G_{11}&=& - \frac{\ud^2}{\ud x^2} + \frac{4 (f + \half)^2 - 1}{4
 x^2} + \la^2 \,x^2+ 2 \la\,(f-\half),\\
\label{Gsd} G_{22}&=& - \frac{\ud^2}{\ud  x^2} + \frac{4 (f-\half)^2 -
1}{4  x^2}+  \la^2 \, x^2+ 2
 \la\,(f+\half).
\enqa

These equations have the  same  structure as  the second order wave equations in AdS space, which follow from a linear Dirac equation with a multiplet structure composed of positive and negative-chirality components~\cite{deTeramond:2013it, deTeramond:2014asa}.  Mapping to light-front physics, one identifies the conformal variable $x$ with $\zeta$, the boost-invariant LF separation of the constituents~\cite{Brodsky:2006uqa}.
\footnote{These equations  are analogous to the LF holographic relation of Eqs.~(5.28) and~(5.29) to Eqs.~(5.32) and~(5.33) in Ref.~\cite{Brodsky:2014yha}.} 
In the case of fermions, the maximal symmetry of AdS was broken in holographic QCD by the introduction of an {\it ad hoc}  Yukawa-like term in the AdS action~\cite{Abidin:2009hr}. This is unnecessary using superconformal algebra.

The operator $G_{22}$ agrees with the LF Hamiltonian of the positive-chirality projection; similarly,  the operator  $G_{11}$ acts on the negative-chirality component. The positive-chirality component  $\psi_+(\zeta) \sim \zeta^{\frac{1}{2} + L} e^{-\la \zeta^2/2} L_n^L(\la \zeta^2)$ has orbital angular momentum  $L_B= f-\half$  and it is the leading twist solution; the  negative-chirality component $\psi_-(\zeta) \sim  \zeta^{\frac{3}{2} + L} e^{-\la \zeta^2/2}  L_n^{L+1}(\la \zeta^2)$ has $L_B+1$. The total nucleon wave function is the plane-wave superposition~\cite{deTeramond:2013it, deTeramond:2014asa}
 \beq \label{Psi}
 \Psi(x^\mu, \zeta) =  e^{i P \cdot x} \left[ \psi^+(\zeta) \tfrac{1}{2}  \left(1 + \gamma_5) u({P}\right) +    \psi^-(\zeta) \tfrac{1}{2}  \left(1 - \gamma_5\right) u({P})\right],
 \enq 
where $u(P)$ is a Dirac spinor of a free nucleon with momentum $P$ in four-dimensional Minkowski space~\cite{deTeramond:2013it, Brodsky:2014yha}. Both components have identical normalization~\cite{deTeramond:2014asa}, and thus the nucleon spin is carried by the LF orbital angular momentum~\cite{Brodsky:2014yha}. The equality of the normalization of the $L=0$ and $L=1$  components   is also predicted by the Skyrme model~\cite{Brodsky:1988ip}.

The operator $G_{11}$ is also the LF Hamiltonian of a meson with angular momentum $J=L_M=f+\half $~\footnote{Compare Eq.~(5.2) with Eq.~(5.5) in Ref{.~\cite{Brodsky:2014yha}}.}. The eigenfunctions of $G_{11}$ and $G_{22}$ are related by the fermionic operators $R_\la$ and $R^\dagger_\la$,  Eqs. \req{R} and \req{Rdag}; these supercharges can be interpreted as operators which transform baryon into meson  wave functions and vice-versa~\cite{Dosch:2015nwa}. The operator $G_{11}$ is thus  the Hamiltonian for mesons, and $G_{22}$  is the Hamiltonian for the positive-chirality component, the leading-twist baryon wave function.

The eigenfunctions of the Hamiltonian $G_{11}$ are  
$\phi_{n,L}(z) \sim \zeta^{1/2 +L} e^{- \la  z^2/2} L_n^L(\la z^2)$, with $L= f+\half$.   Using the relations
$n L_n^\nu(x) = (n + \nu)  L_{n-1}^\nu(x) - x  L_{n-1}^{\nu+1}(x)$ and $  L_n^{\nu-1}(x) =  L_n^\nu(x) -  L_{n-1}^\nu(x)$  between the associated Laguerre polynomials we find 
\beq \label{Rdagphi} 
R_\la ^\dagger \vert \phi^M_{n,L} \rangle = 2 \sqrt \la (n+ L)^{1/2}  \vert \phi^B_{n,L-1} \rangle,
\enq
where 
\beq 
|\phi^M_{n,L} \rangle=\left(\begin{array}{c} \phi_{n,L}\\ 0 \end{array}\right), \quad \quad
|\phi^B_{n,L-1} \rangle=\left(\begin{array}{c}0\\  \phi_{n,L-1}\end{array}\right). 
\enq 
This shows explicitly  the remarkable relation $L_M = L_B + 1$ which identifies the  orbital angular momenta of the mesons  with their baryon superpartners with identical mass~\cite{Dosch:2015nwa}.

The relation $L_B=f-\half$ shows that $f$ must be positive for baryons, in accordance with the requirement that the superconformal potential $\frac{f}{x}$ in \req{q} be bounded from below~\cite{Fubini:1984hf, deAlfaro:1976je}.  However, for mesons,  the negative value $f=-\half$ leads to angular momentum $L_M=0$, which is allowed  and is consistent with the Hamiltonian $G_{11}$ for mesons~\footnote{In LFHQCD the lowest possible value $L_M=0$ corresponds to the lowest possible value for the AdS$_5$ mass allowed by the Breitenlohner-Freedman stability bound~\cite{Breitenlohner:1982jf}.}. We can therefore regard  Eq.~\req{Gsc} as an extension of the supersymmetric theory with $f>0$ to the negative value $f=-\half$ for mesons. It is clear from Eq. \req{Rdagphi} that the fermion operator $R^\dagger$ annihilates the lowest state corresponding to $n = L = 0$, $R^\dagger  \vert \phi_{n = 0, L = 0} \rangle = 0$,  in accordance with  Eq.~\req{anni}. Thus the pion has a special role in the superconformal  approach to hadronic physics as a unique state of zero mass~\cite{Dosch:2015nwa}.  It also follows from Eq.~\req{Rdagphi} that meson states with $n>0$ and $L=0$, also corresponding to the marginal value  $f=-\half$, are not annihilated by $R^\dagger$. These  states, however, are connected to  unphysical fermion  states with $L=-1$.  These spurious states are eigenstates of the Hamiltonian $G$, Eq. \req{G}, with positive eigenvalues  and their presence seems unavoidable in the supersymmetric construction, since each state with eigenvalue different from zero should have a partner, as dictated by the index theorem~\cite{Witten:1981nf}.

The situation is completely analogous to the case where  conformal symmetry is explicitly and strongly  broken by heavy quark masses.  In this case the superpotential  is no longer  constrained by conformal symmetry and it is basically unknown, but the meson-baryon supersymmetry still holds~\cite{Dosch:2015bca}.  
In particular,  the   $L = 0$ meson states have no supersymmetric baryon partner since they would correspond to  unphysical $L = -1$ states.

\section{Supersymmetric Aspects of Hadron Physics~\cite{Brodsky:2016yod}}

As discussed in the previous section, the light-front holographic results can be extended~\cite{deTeramond:2014asa,Dosch:2015nwa,Dosch:2015bca} to effective QCD light-front equations for both mesons and baryons by using the generalized supercharges of superconformal algebra~\cite{Fubini:1984hf}.   In effect the baryons are color-singlet bound-states of color-triplet quarks and $\bar 3_C$ $[qq]$ diquarks.  
This novel approach to hadron physics not only allows a treatment of nucleons which is  analogous to that of mesons, but  it also captures  the essential properties  of the  confinement dynamics of light hadrons  and provides a theoretical foundation for the observed similarities between mesons and baryons. The superconformal algebraic approach can be extended to include the spin-spin interactions of the constituents and the contribution to the hadron spectrum from quark masses.

\begin{figure}
 \begin{center}
\includegraphics[height=11cm,width=15cm]{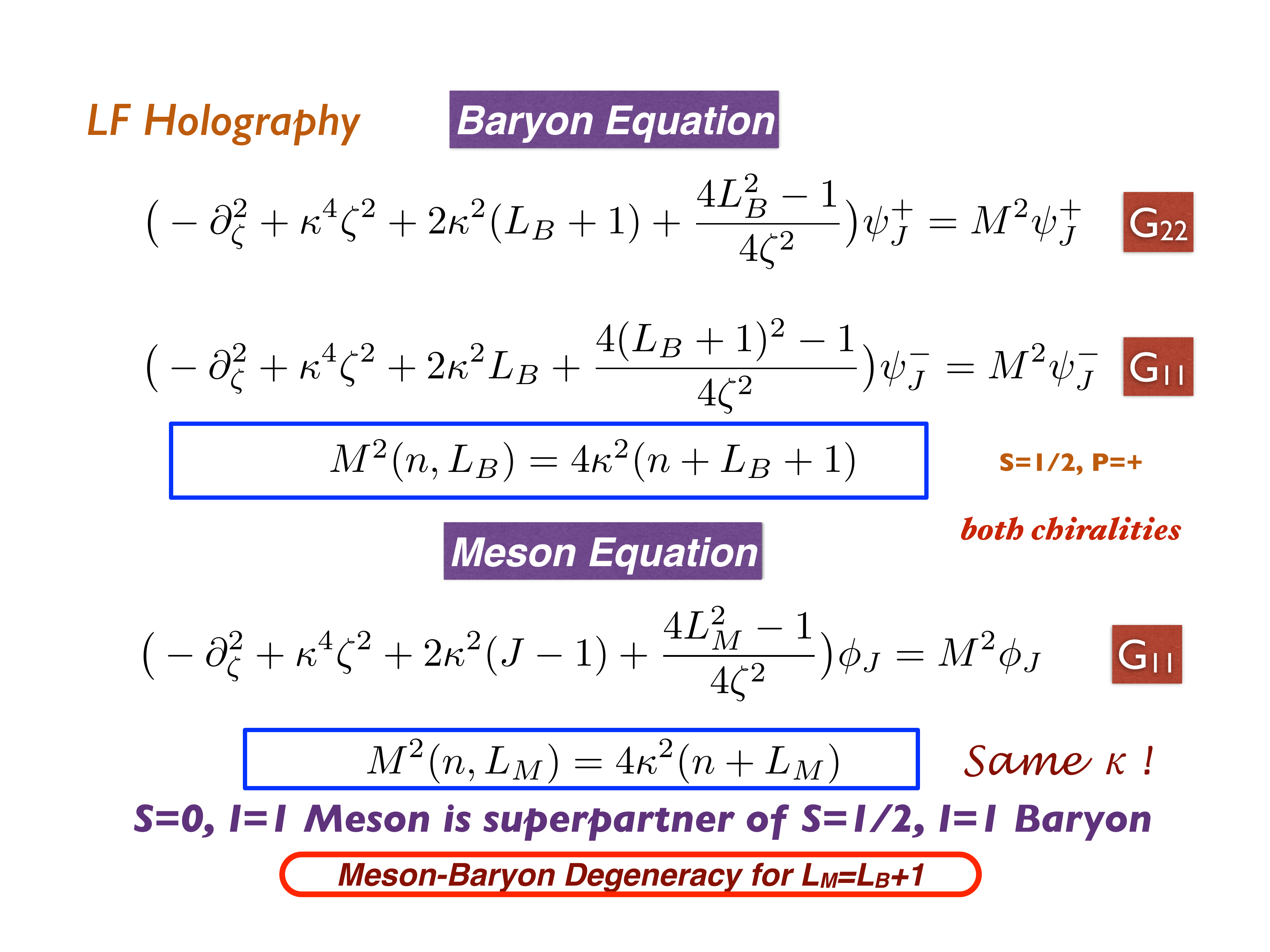}
\end{center}
\caption{ The LF Schr\"odinger equations for baryons and mesons for zero quark mass derived from the  superconformal algebraic approach to hadron physics.
The $\psi^\pm$  are the baryon quark-diquark LFWFs where the quark spin $S^z_q=\pm 1/2$ is parallel or antiparallel to the baryon spin $J^z=\pm 1/2$.   The meson and baryon equations are identical if one identifies a meson with internal orbital angular momentum $L_M$ with its superpartner baryon where  $L_M = L_B + 1.$
See Refs.~\cite{deTeramond:2014asa,Dosch:2015nwa,Dosch:2015bca}.
}
\label{FigsJlabProcFig3.pdf}
\end{figure}

The supercharges connect the baryon and meson spectra  and their Regge trajectories to each other in a remarkable manner: each meson has internal  angular momentum one unit higher than its superpartner baryon  $L_M = L_B+1.$  See  Fig. \ref {FigsJlabProcFig3.pdf}.   Only one mass parameter $\kappa =  \sqrt \lambda$ again appears; it sets the confinement and the hadron mass scale in the  chiral limit, as well as  the length scale which underlies hadron structure.  ``Light-Front Holography"  not only predicts meson and baryon  spectroscopy  successfully, but also hadron dynamics, including  vector meson electroproduction,  hadronic light-front wavefunctions, distribution amplitudes, form factors, and valence structure functions.  
The LF Schr\"odinger Equations for baryons and mesons derived from superconformal algebra  are shown  in Fig. \ref{FigsJlabProcFig3.pdf}.
In effect the baryons on the proton (Delta) trajectory are bound states of a quark with color $3_C$ and a scalar (vector)  diquark with color $\bar 3_C$. 
The proton eigenstate labeled $\psi^+$ (parallel quark and baryon spins) and $\psi^-$ (anti parallel quark and baryon spins)  have equal Fock state probabilities, 
a specific form of  chirality invariance.  The static properties of the nucleons are discussed in Ref.~\cite{Liu:2015jna}.

The comparison between the predicted meson and baryon masses of the meson and baryon Regge trajectories is shown in Fig. \ref{SUSYduo}.
Superconformal algebra  predicts that the bosonic meson and fermionic baryon masses are equal if one identifies each meson with internal orbital angular momentum $L_M$ with its superpartner baryon with $L_B = L_M-1.$      Since $ | L_B - L_M | =1,$    the meson and baryon  superpartners thus have have the same parity as well as the same twist. Notice that the twist  $ 2+ L_M = 3 + L_B$ of the interpolating operators for the meson and baryon superpartners are the same.    Superconformal algebra also predicts 
relations between the LFWFs of the superpartners.   The predicted identity of meson and baryon timelike form factors can be tested in $e^+ e^- \to H \bar H^\prime $ reactions. 

The  dynamical AdS/QCD soft-wall predictions for the meson and nucleon form factors are compared with experiment in ref.~\cite{Brodsky:2007hb,Brodsky:2014yha}.   
This approach has also been extended to the charge, magnetic, and quadruple elastic form factors of the deuteron by Gutsche, et al. ~\cite{Gutsche:2015xva}. The predictions are in remarkable agreement with measurements.~\cite{Abbott:2000ak,Holt:2012gg}

\subsection{Holographic embedding and the spin-spin interaction~\cite{Brodsky:2016yod}}

The pion and nucleon trajectories can be consistently described by the superconformal algebraic structure mapped to the light front~\cite{Dosch:2015nwa}.  A fundamental prediction is a  massless pion in the limit of zero quark masses. However, there remains a lingering question for the $\rh$ and $\Delta$ trajectories:  in the case of light-front holographic QCD,  it was found necessary  to introduce the concept of half-integer twist~\cite{Brodsky:2014yha, Dosch:2015nwa}  in order to describe the $\Delta$ trajectory.
For states with $J= L_M+s$, where $s$ is the total quark spin, the spin interaction follows from the holographic embedding of the bound-state equations~\cite{Brodsky:2014yha}; it is not determined by the superconformal construction. This amounts to the modification of the meson Hamiltonian $G_{11} \to G_{11} + 2 \la\,  s$.   In  order to preserve supersymmetry,  one must add the same term to the baryon Hamiltonian $G_{22}$. The  resulting supersymmetric Hamiltonian for mesons and baryons in the chiral limit  is therefore
\beq \label{GS}
G_S=  \{R_\la,R^\dagger_\la\}  +  2 \la \,  s \, \mathbf{I}.
\enq
For mesons  $s$ is the total internal quark spin of a meson. The identification of baryons  as bound states of a quark  and a  diquark cluster provides a  satisfactory interpretation of the supersymmetric implementation: in  this  case we can identify $s$ with the spin of the diquark cluster. The spin of the diquark cluster of the $\Delta$ trajectory and the nucleon family with total quark spin $\frac{3}{2}$ must be  $s=1$: it is the natural superpartner of the $\rho$ trajectory. For the nucleon family with total quark spin $\half$, the cluster  is, in general, a superposition  of spin  $s=0$ and $s=1$. Since the nucleon trajectory is the natural partner of the $\pi$ trajectory, we have to choose the cluster spin $s=0$ to maintain supersymmetry. In general, we  take $s$ as the smallest possible value compatible with the quantum numbers of the  hadrons and  the Pauli principle. This procedure reproduces the agreement with the empirical baryon spectrum obtained in our previous treatments without  the unsatisfactory feature of introducing half-integer twist;  all twists and orbital angular momenta are integers.

In the case of mesons, the lowest mass state of the vector meson family, the $I=1$ $\rho$ (or the $I=0$ $\omega$ meson) is  annihilated by the fermion operator $R^\dagger$, and it has no baryon superpartner.  This is possible, even though the $\rho$ is a massive particle in the limit of zero quark masses, since the effect of the spin term $2 \la s$ in the new Hamiltonian  is an overall shift of the mass scale without a modification of the LF wavefunction. The action of the fermion operator is thus the same as for the pseudoscalar meson family.

To summarize: The meson wave function $ \phi_M(L_M)$, with LF orbital angular momentum $L_M$ and quark spin $s$, and the positive-chirality (leading-twist) component wave function $\psi_{B+}$  of a baryon, with cluster spin $s$ and orbital angular momentum $L_B=L_M-1$, are part of the supermultiplet
\beq \label{MBplet}
\vert \phi_H\rangle = \left( \begin{array}{c} \phi_M(L_M) \\ \psi_{B+}(L_B=L_M-1) \end{array} \right),
\enq
with equal mass.  The supercharge $R_\lambda^\dagger$ acts on the multiplet \req{MBplet} and transforms the meson wave function into the corresponding baryon wave function. The meson and baryon mass spectra resulting from the Hamiltonian \req{GS} are given by the simple formul\ae:
\beqa \label{hadchiral}  
\mbox{Mesons } && M^2 = 4 \la (n+L_M)+ 2 \la \,  s ,\\
\label{mesfin}\mbox{Baryons} && M^2=4 \la(n+L_B+1) + 2 \la \,  s .
\enqa
As discussed below Eq. \req{GS},  $s$ is the internal quark spin for mesons  and the lowest possible value of the cluster spin for baryons  .

We are working in a semiclassical approximation;  therefore hadronic states are described by wave functions in the Hilbert space $\cL_2(R_1)$,  where the  variable is the boost-invariant light-front transverse separation $\zeta$. The generators of the symmetries are operators in that Hilbert space. Since the wave functions (and spectra) are equal for hadrons and anti-hadrons, the superpartner of the meson is a baryon as well as the corresponding antibaryon. 

In order to interpret these results for hadron physics,  we assume that the constituents of  mesons and baryons are quarks or antiquarks with the well-known quantum numbers. Thus the fermion operator $R_\lambda^\dagger$ is interpreted as the transformation operator of a  single constituent (quark or antiquark)  into an anti-constituent cluster  in the conjugate color representation.

\subsection{Tetraquarks~\cite{Brodsky:2016yod}}

\begin{figure}
\setlength{\unitlength}{1mm}
\begin{center}
\begin{picture}(60,30)(15,0)
\put(20,30){\circle{20}} \put(20,33){\circle*{2}}
\put(20,27){\circle{2}}
 \put(50,30){\circle{20}}
\put(48,33){\circle{2}} \put(52,33){\circle{2}}
\put(50,27){\circle{2}}
 \put(14,19){$\phi_M,\;L_B+1$}
\put(43,19){$\psi_{B+},\;L_B$}
 \put(22,33){\vector(1,0) {24}}
\put(32,35){$R_\lambda^\dagger$}

 \put(50,7){\circle{20}}
\put(48,10){\circle{2}} \put(52,10){\circle{2}}
\put(50,4){\circle{2}} \put(43,-4){$\psi_{B-}, \;L_B+1$}
\put(80,7){\circle{20}} \put(78,10){\circle{2}}
\put(82,10){\circle{2}} \put(78,4){\circle*{2}}
\put(82,4){\circle*{2}} \put(73,-4){$\phi_T,\;L_B$}
\put(52,4){\vector(1,0) {24}} \put(62,6){$R_\lambda^\dagger$}
\end{picture}
\end{center}
\caption{\label{tetra} \small The supersymmetric quadruplet $\{\phi_M, \psi_{B+}, \psi_{B-},\phi_T\}$. Open circles represent quarks, full circles antiquarks. The tetraquark has the same mass as its baryon partner in the multiplet. Notice that the LF angular momentum of the negative-chirality component wave function of a baryon $\psi_{B-}$ is one unit higher than that of the positive-chirality (leading-twist) component $\psi_{B+}$.}
\end{figure}
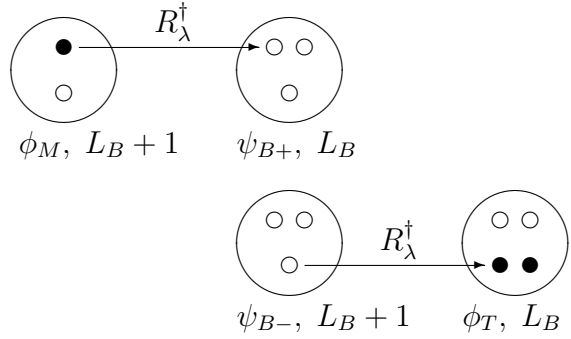

The supersymmetric states introduced in the previous section do not form a complete supermultiplet since the negative-chirality component wave function of the baryon has not been assigned as yet to its supersymmetric partner.   We can complete the supersymmetric multiplet by applying the fermion operator $R_\la^\dagger$ to the negative-chirality component baryon wave function and thus obtain a new bosonic state.   As noted above, the operator $R_\la^\dagger$ can be interpreted as transforming a constituent into a two-anticonstituent cluster.   Therefore  the operator $R_\la^\dagger$ applied to the negative-chirality component of a baryon generates a tetraquark wave function,  $\phi_T= R_\la^\dagger\, \psi_{B-}$,  a bound state of a diquark and an anti-diquark cluster  as depicted in Fig.~\ref{tetra}.

The negative-chirality component of a baryon, $\psi_{B-}$, has LF angular momentum $L_B+1$ if its positive-chirality component partner has LF angular momentum $L_B$.  Since $R_\la^\dagger$  lowers the angular momentum by one unit, the angular momentum of the corresponding tetraquark is  equal to that of the positive-chirality component of the baryon, $L_T=L_B$. The complete quadruplet supersymmetric representation thus consists of two fermion wave functions, namely the positive and negative-chirality components of the baryon spinor wavefunction $\psi_{B+}$ and $\psi_{B-}$, plus two bosonic wave functions, namely the meson $\phi_B$ and the tetraquark $\phi_T$.  These states can be arranged as a $2 \times2$ matrix: \beq
 \left(\begin{array}{cc}
\phi_M{(L_M = L_B+1)} &\psi_{B-}{(L_B+ 1)}\\
\psi_{B+}{(L_B)} &\phi_T{(L_T = L_B)} 
\end{array} 
\right),
\enq 
on which the symmetry generators~\req{susy-extended} and the Hamiltonian \req{GS} operate~\footnote{It is interesting to note that in Ref.~\cite{Catto:1984wi} mesons, baryons and tetraquarks  are also hadronic states within  the same multiplet.}.

According to  Eq.~\req{GS} the total quark spin of all states must be the same. Furthermore we have to take into account that the diquark as a two-fermion state has to be totally antisymmetric. The colour indices are antisymmetric and therefore the spin and isospin of a cluster of two light quarks ($u,\,d$) are correlated. Quark spin $s=0$ goes together with isospin $I=0$,  and $s=1$ entails  $I=1$. The resulting cluster configurations for several families of baryons and their tetraquark partners are displayed in Table \ref{tetratable}.

{\renewcommand{\arraystretch}{1.1}  
\begin{table}[h]  
\begin{center}
\begin{tabular}{| l | ccc|ccc|}
\hline
& \multicolumn{3}{c|}{Baryon} & \multicolumn{3}{c|}{Tetraquark}\\
& & $s$ & $I$ & &$s$ &$I$ \\
\hline
N-& $q$ & $\half$ & $\half$ & $(\bar q \bar q)$ & 0 & 0 \\
fam. & $(qq)$ & 0 & 0 & $(qq)$ & 0 & 0 \\
\hline
$\De$-& $q$ & $\half$ & $\half$ & $(\bar q \bar q)$ & 0 & 1 \\
fam. & $(qq)$ & 1 & 1 & $(qq)$ & 1 & 1 \\
\hline
$\La$-& $s$ & $\half$ & $0$ & $(\bar s \bar q)$ & 0 & $\half$ \\
fam. & $(qq)$ & 0 & 0 & $(qq)$ & 0 & 0 \\
\hline
$\Si$-& $q$ & $\half$ & $\half$ & $(\bar q \bar q)$ & 0 & 0\\
fam. & $(sq)$ & 0 & $\half$ & $(sq)$ & 0 & $\half$ \\
\hline
$\Si^*$-& $s$ & $\half$ & $0$ & $(\bar s \bar q)$ & 0 & $\half$ \\
fam. & $(qq)$ & 1 & 1 & $(qq)$ & 1 & 1 \\
\hline
$\Xi$-& $s$ & $\half$ & $0$ & $(\bar s \bar q)$ & 0 & $\half$ \\
fam. & $(sq)$ & 0 & $\half$ & $(sq)$ & 0 & $\half$ \\
\hline
$\Xi^*$-& $s$ & $\half$ & $0$ & $(\bar s \bar q)$ & 0 & $\half$ \\
fam. & $(sq)$ & 1 & $\half$ & $(sq)$ & 1 & $\half$ \\
\hline
\end{tabular}
\end{center}
\begin{footnote}
 ~~Quantum numbers of the constituents and constituent clusters of different baryon families and their supersymmetric tetraquark partners.
\end{footnote}
\label{tetratable}
\end{table}

The quantum numbers of the tetraquark itself can be easily calculated from the ones of the two constituent clusters.  Since the relative angular momentum of the two diquarks in the tetraquark is equal to the angular momentum $L_B$ of the positive-chirality component of the baryon, and since the tetraquark consists of an even number of antiquarks, its parity is $(-1)^{L_B}$.

The leading-twist component of the nucleon has $L_B=0, s=0$. Thus its tetraquark partner consists of a diquark and anti-diquark, both with $s=0$; therefore its isospin is $ I=0$. The parity must be $P=+$,  since it has $L=0$ and it consists of two particles and two antiparticles.  A candidate for such a state is the $f_0(980)$. For the partner of the $\De$ resonance we must have $s=1$, it therefore consists of a diquark with  $I=1,\; s=1$ and an anti-diquark with $I=0,\;s=0$. The resulting quantum numbers are $I=1, s=1$ and $P=+$;  the $a_1(1260)$ is a candidate.   The first $L$ excitation of the nucleon is the $N^{3/2-}(1520)$ and $N^{1/2-}(1535)$ pair.  Its tetraquark partner consists of two $I=0,  s=0$ clusters, and thus its quantum numbers are $I=0, J^P =0,1^-$;  candidates are the $\omega(1420)$ and $\omega(1650)$ -- or the mixing of these two states.

\subsection{Inclusion of quark masses and comparison with experiment~\cite{Brodsky:2016yod}}

We have shown in~\cite{Brodsky:2014yha} that the natural way to include light quark masses in the hadron mass spectrum is to leave the LF potential unchanged  as a first approximation and include the additional term of the invariant mass $ \Delta m^2 = \sum_{i=1}^n \frac{m_i^2}{x_i}$ in the LF kinetic energy.  In fact, the ${m^2_q\over x_q}$ contribution to the LF Hamiltonian can be derived from the  spin-flip Yukawa interaction of a confined quark with the background LF zero-mode 
Higgs field of the Standard Model~\cite{Srivastava:2002mw}.

The resulting LF wave function is then  modified by  the factor $e^{-\frac{1}{2\la} {\Delta m^2}}$, thus providing a relativistically invariant form  for the hadronic wave functions.  The effect of the nonzero quark masses for the squared hadron masses is then given by the expectation value of $ \Delta m^2$ evaluated using the modified wave functions. This prescription leads to the quadratic mass correction
\beq \label{DeM}
\De M^2[m_1,\cdots,m_n]=\frac{\lambda^2}{F}\,\frac{\ud F}{\ud\lambda},
\enq
with
$F[\lambda]=\int_0^1\cdots \int \ud x_1  \cdots \,\ud x_n
\,  e^{-\frac{1}{\la}\left(\sum_{i=1}^n
\frac{m_i^2}{x_i}\right)}\de(\sum_{i=1}^n x_i-1)$.

The resulting expressions for the squared masses of all light mesons and baryons are:
\beqa \label{mesfin}  
\mbox{Mesons} && M^2 = 4 \la (n+L)+ 2 \la \, s + \De M^2[m_1,m_2] ,\\
\label{barfin}\mbox{Baryons} && M^2=4 \la(n+L+1) + 2 \la \, s + \De M^2[m_1,m_2,m_3] ,
\enqa 
where the different values of the mass corrections within the supermultiplet break supersymmetry explicitly. For the tetraquark the mass formula is the same as for the baryon except for the quark mass correction  $\De M^2[m_1,m_2,m_3,m_4]$ given by Eq. \req{DeM}.

The pion mass of $\sim 0.140$ GeV is obtained if the non-strange light-quark mass is $m=0.045$ GeV~\cite{Brodsky:2014yha}.   In the case of the  $K$-meson,  the resulting value for the strange quark mass is $m_s=0.357$ GeV~\cite{Brodsky:2014yha}.    The trajectories of $K$, $K^*$ and $\phi$-mesons can then be readily calculated. (The predictions  are compared with experiment in Ref.~\cite{Brodsky:2014yha}. )  In Eq.~\req{DeM} the values of $x_i$ for the quarks are assumed to be uncorrelated. If one instead assumes maximal correlations in the cluster, i.e. $x_2=x_3$, this affects the final result by less than 1~\% for light quarks and less than 2~\% for the $\Omega^-$  which has three strange quarks. Therefore,  the  previously obtained agreement with the data~\cite{deTeramond:2014asa}  for the baryon spectra is hardly affected.   

One can fit the value of the fundamental mass parameter $\sqrt{\la}$ for each meson and baryon Regge  trajectory separately using Eqs.~\req{mesfin} and~\req{barfin} . The results are displayed in Fig.~\ref{slope}.   The best fit gives 
$\sqrt \lambda = 0.52$ GeV as the characteristic mass scale of QCD.

\begin{figure}
\begin{center}
\includegraphics[height=6.5cm,width=7cm]{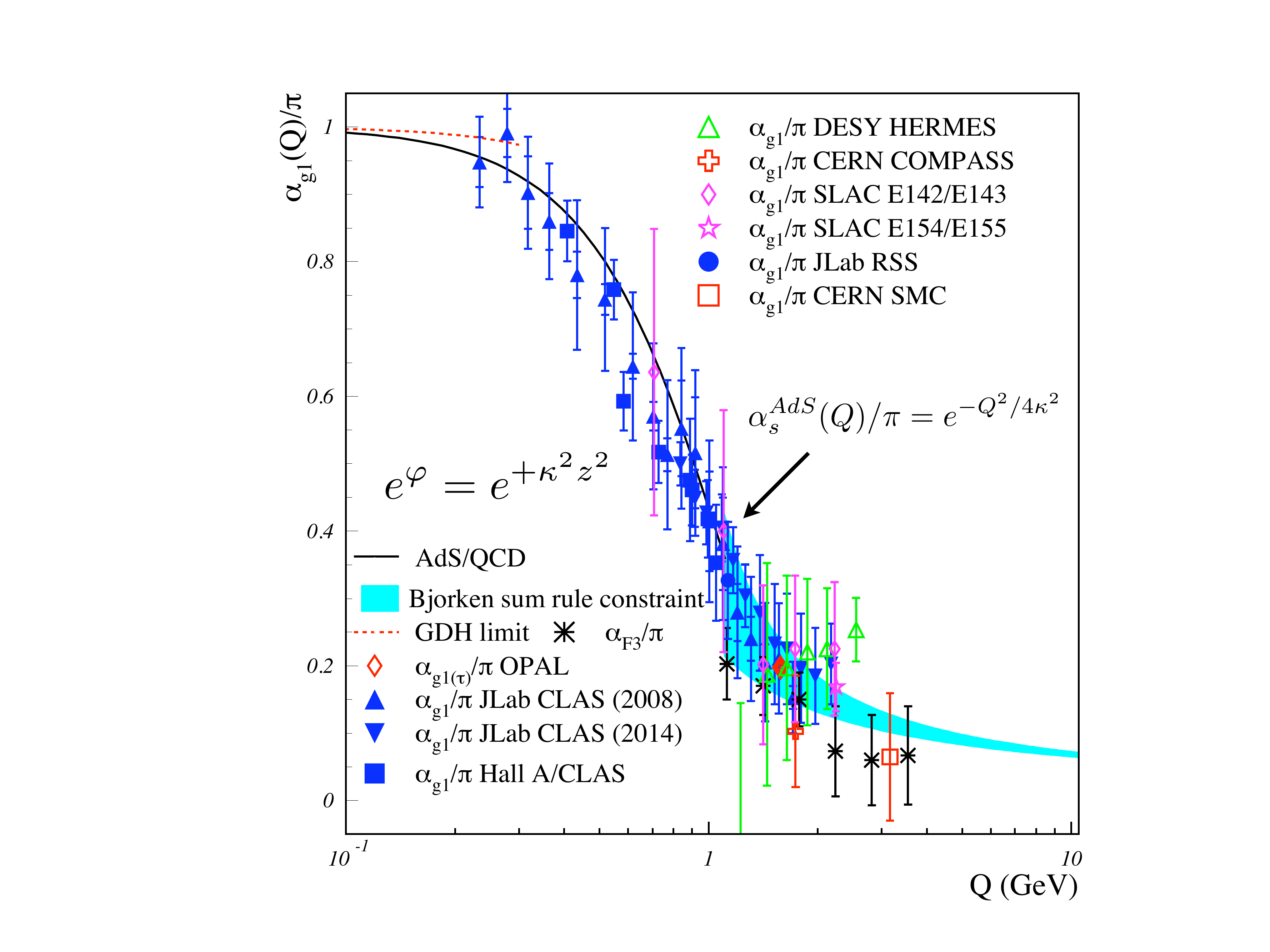}
\includegraphics[height=6.5cm,width=7cm]{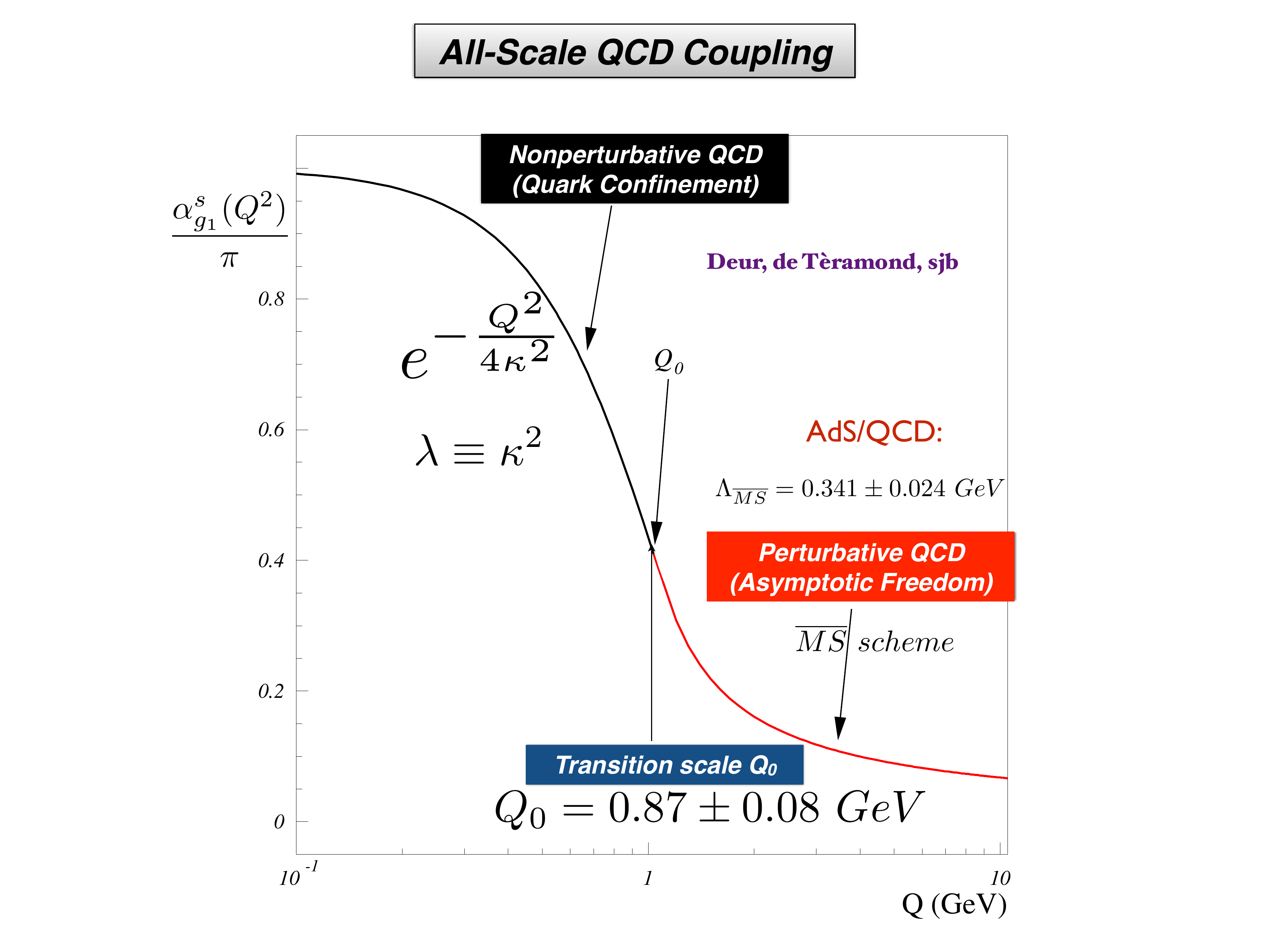}
\end{center}
\caption{
(A) Comparison of the predicted nonpertubative coupling, based on  the dilaton $\exp{(+\kappa^2 z^2)}$ modification of the AdS$_5$ metric, with measurements of the effective charge $\alpha^s_{g_1}(Q^2)$  
defined from the Bjorken sum rule.
(B)  Prediction from LF Holography and pQCD for the QCD running coupling $\alpha^s_{g_1}(Q^2)$ at all scales.   The magnitude and derivative of the perturbative and nonperturbative coupling are matched at the scale $Q_0$.  This matching connects the perturbative scale 
$\Lambda_{\overline{MS}}$ in the ${\overline{MS}}$  scheme to the nonpertubative scale $\kappa$ which underlies the hadron mass scale. 
See Ref.~\cite{Brodsky:2014jia}. }
\label{CouplingMatching}
\end{figure} 

\section {The QCD Coupling at all Scales} 

The QCD running coupling $\alpha_s(Q^2)$
sets the strength of  the interactions of quarks and gluons as a function of the momentum transfer $Q$.
The dependence of the coupling
$Q^2$ is needed to describe hadronic interactions at 
both long and short distances. 
The QCD running coupling can be defined~\cite{Grunberg:1980ja} at all momentum scales from a perturbatively calculable observable, such as the coupling $\alpha^s_{g_1}(Q^2)$, which is defined from measurements of the Bjorken sum rule.   At high momentum transfer, such ``effective charges"  satisfy asymptotic freedom, obey the usual pQCD renormalization group equations, and can be related to each other without scale ambiguity 
by commensurate scale relations~\cite{Brodsky:1994eh}.  

The dilaton  $e^{+\kappa^2 z^2}$ soft-wall modification of the AdS$_5$ metric, together with LF holography, predicts the functional behavior of the running coupling
in the small $Q^2$ domain~\cite{Brodsky:2010ur}: 
${\alpha^s_{g_1}(Q^2) = 
\pi   e^{- Q^2 /4 \kappa^2 }}. $ 
Measurements of  $\alpha^s_{g_1}(Q^2)$ are remarkably consistent~\cite{Deur:2005cf}  with this predicted Gaussian form; the best fit gives $\kappa= 0.513 \pm 0.007~GeV$.   
See Fig.~\ref{CouplingMatching}(A).

In a collaboration with Alexandre Deur~\cite{Brodsky:2010ur,Deur:2014qfa,Brodsky:2014jia}, we have shown how the parameter $\kappa = \sqrt \lambda$,  which   determines the mass scale of  hadrons in the zero quark mass limit, can be connected to the  mass scale $\Lambda_s$  controlling the evolution of the perturbative QCD coupling.  The high momentum transfer dependence  of the coupling $\alpha_{g1}(Q^2)$ is  specified by  pQCD and its renormalization group equation.  The 
matching of the high and low momentum transfer regimes  of $\alpha_{g1}(Q^2)$ -- both its value and its slope -- then determines the scale $Q_0$ setting the interface between perturbative and nonperturbative hadron dynamics.  This connection can be done for any choice of renormalization scheme, such as the $\overline{MS}$ scheme,
as seen in   Fig.~\ref{CouplingMatching}(B).
The result of this perturbative/nonperturbative matching is an effective QCD coupling  defined at all momenta.   A comprehensive review the QCD running coupling is given in ref.~\cite{Deur:2016tte}.

The predicted value of $\Lambda_{\overline{MS}} =  0.341 \pm 0.024~GeV$ from this analysis agrees well the measured value~\cite{Agashe:2014kda}  
$\Lambda_{\overline{MS}} = 0.339 \pm 0.016~GeV$.  Conversely, we can predict the value of  $\kappa = \sqrt \lambda$ and the hadron mass spectrum for light quarks using the experimental value of $\Lambda_{\overline{MS}} $  as the sole input parameter.
Thus these results, combined with the AdS/QCD superconformal predictions for hadron spectroscopy, allow us to compute hadron masses in terms of $\Lambda_{\overline{MS}}$:
$m_p =   2 \kappa = 3.21~ \Lambda_{\overline{MS}},~ m_\rho = { \sqrt  2}\kappa = 2.2 ~ \Lambda_{\overline{ MS} }, $ and $m_p = \sqrt 2 m_\rho, $ meeting a challenge proposed by Zee~\cite{Zee:2003mt}.
The pion is predicted to be massless for $m_q=0$  consistent with chiral theory.

The value of $Q_0$ can be used to set the factorization scale for DGLAP evolution of hadronic structure functions and the ERBL evolution of distribution amplitudes.
The dependence of $Q_0$ on the choice of the  effective charge used to define the running coupling and the renormalization scheme used to compute its behavior in the perturbative regime has also been determined~\cite{Brodsky:2010ur,Deur:2014qfa,Brodsky:2014jia},.

\section{Hadronization at the Amplitude Level and Other New Directions}

\begin{itemize}

\item
The new insights into color confinement given by AdS/QCD suggest that one could compute hadronization at  amplitude level~\cite{Brodsky:2009dr} using LF time-ordered perturbation theory, but including the confinement interaction.  For example, if one computes $e^+ e^- \to q \bar q \to q \bar q g \cdots$, the quarks and gluons only appear in intermediate states, and only hadrons can be produced.  LF perturbation theory 
provides a remarkably efficient method for the calculation of multi-gluon amplitudes~\cite{Cruz-Santiago:2015nxa}. 

\item
The eigensolutions of the AdS/QCD LF Hamiltonian can used to form an ortho-normal basis for diagonalizing the complete QCD LF Hamiltonian.  This method, ``basis light-front quantization"~\cite{Vary:2009gt}  is expected to be more efficient than the DLCQ method~\cite{Pauli:1985pv} for obtaining QCD 3+1  solutions.

\item
The $\kappa^4 \zeta^2$ confinement interaction between a $q$ and $\bar q$ will induce a $\kappa^4/s^2$ correction to $R_{e^+ e^-}$, replacing the $1/ s^2$ signal usually attributed to a vacuum gluon condensate.  

\item
The kinematic condition that all $k^+ = k^0+ k^3$ are positive and conserved precludes QCD condensate contributions to the $P^+=0$ LF vacuum state, which by definition is the causal, frame-independent lowest invariant mass eigenstate of the LF Hamiltonian~\cite{Brodsky:2009zd,Brodsky:2012ku}. 

\item
It is interesting to note that the contribution of the {\it `H'} diagram to $Q \bar Q $ scattering is IR divergent as the transverse separation between the $Q$  
and the $\bar Q$ increases~\cite{Smirnov:2009fh}.  This is a signal that pQCD is inconsistent without color confinement.  The sum of such diagrams could sum to the confinement potential $\kappa^4 \zeta^2 $ dictated by the dAFF principle that the action remains conformally invariant despite the mass scale in the Hamiltonian.

\item
The AdS/QCD confinement potential also has impact on the physics underlying ``ridge" product in high multiplicity hadron-hadron collisions~\cite{Bjorken:2013boa}.  The color confining interaction  binding the colored constituents is analogous to a gluonic string or flux tube.
In the case of $pp $ collisions, if the  flux tubes connecting the quark and di-quark cluster of each proton are aligned, one will produce maximum multiplicity and ridge-like structures. Similarly, 
In the case of $e p \to e^\prime X$, one can consider the collisions of the confining  QCD flux tube appearing between the $q$ and $\bar q$  of the virtual photon with the flux tube between the quark and diquark of the proton.   Since the $q \bar q $ plane tends to be aligned with the scattered electron's plane, the resulting ``ridge"  of hadronic multiplicity produced from the $\gamma^* p$ collision will also tend to be aligned with the scattering plane of the scattered electron.  The virtual photon's flux tube will also depend on the photon virtuality $Q^2$, as well as the flavor of the produced pair arising from $\gamma^* \to q \bar q$.  In the case of high energy ultra-peripheral p p collisions, one can control the  produced  hadron multiplicity and ridge geometry using the scattered protons' planes. The resulting dynamics~\cite{Brodsky:2014hia} is  a natural extension of the flux-tube collision description of the ridge produced in $p-p$ collisions~\cite{Bjorken:2013boa}.

\end{itemize}

\section{Elimination of  Renormalization and Factorization Scale Ambiguities}

The ``Principle of Maximum Conformality", (PMC)~\cite{Wu:2013ei} systematically eliminates the renormalization scale ambiguity in perturbative QCD calculations, order-by-order.    The PMC predictions are also insensitive to the choice of the initial renormalization scale $\mu_0.$
The PMC sums all of the non-conformal terms associated with the QCD $\beta$ function into the scales of the coupling at each order in pQCD, systematically extending the BLM procedure~\cite{Brodsky:1982gc} to all orders. 
The resulting  conformal series is free of renormalon resummation problems.  The number
of active flavors $n_f$ in the QCD $\beta$ function is also
correctly determined at each order. 
 
The $R_\delta$ scheme -- a generalization  of t'Hooft's  dimensional regularization systematically  identifies the nonconformal $\beta$ contributions to any perturbative QCD series, thus allowing the automatic implementation of the PMC procedure~\cite{Mojaza:2012mf}.     
 The resulting scale-fixed predictions for physical observables using
the PMC are {\it  independent of
the choice of renormalization scheme} --  a key requirement of 
renormalization group invariance.    
A related approach is given in Refs.~\cite{Kataev:2014zha,Kataev:2014jba,Kataev:2014zwa}.
 
The elimination of renormalization scale ambiguities greatly increases the precision, convergence, and reliability of pQCD predictions.  
For example, PMC scale-setting has been applied to the pQCD prediction for $t \bar t$ pair production at the LHC,  where subtle aspects of the renormalization scale of the three-gluon vertex and multi-gluon amplitudes, as well as  large radiative corrections to heavy quarks at threshold play a crucial role.  
The large discrepancy of pQCD predictions with  the $t \bar t$  forward-backward asymmetry measured at the Tevatron is significantly reduced from 
$3~\sigma$ to approximately $ 1~\sigma$~\cite{Brodsky:2012rj,Brodsky:2012sz,Wu:2015rga,Wang:2015lna}.  

The use of  the scale $Q_0$ discussed in the previous section to  resolve  the factorization scale uncertainty in structure functions and fragmentation functions,  in combination with the PMC for  setting the  renormalization scales,  can 
greatly improve the precision of pQCD predictions for collider phenomenology.

\section{Conclusions}

Inspired by the correspondence of classical gravitational theory  in 5-dimensional AdS space with superconformal quantum field theory in physical 4-dimensional space-time, as originally proposed by Maldacena~\cite{Maldacena:1997re},  we have arrived at a novel holographic application of supersymmetric quantum mechanics to light-front quantized Hamiltonian theory in physical space-time. The resulting superconformal algebra, which is the basis of our semiclassical theory, not only determines the breaking of the maximal symmetry of the dual gravitational theory, but it also provides the form of the frame-independent  color-confining light-front potential in the semiclassical theory.  

We  have derived a semiclassical  light-front relativistic Hamiltonian for hadron physics based on superconformal algebra and its holographic embedding, which includes a spin-spin interaction between the hadronic constituents.   This extension of our previous results provides a remarkably simple, universal  and consistent description of the  light-hadron spectroscopy and their light-front wavefunctions.   We also predict the existence of tetraquarks which are degenerate with baryons with the same angular momentum. The tetraquarks are bound states of the same confined color-triplet diquark and anti-diquark clusters which account for baryon spectroscopy; they are required to complete the supermultiplet structure  predicted by the superconformal algebra.

Remarkably, supersymmetric quantum mechanics, together with light-front holography, account for many important  features of hadron physics, such as the approximatively linear Regge trajectories (including the daughter trajectories) with nearly equal slopes for all mesons and baryons in both $L$  and $n$.  One finds remarkable supersymmetric aspects of hadron physics, connecting the masses of mesons and their superpartner baryons which are related by $L_M=L_B+1$.  The predicted spectroscopy for the meson and baryon superpartners agree with the data up to  an average absolute deviation of about 10 \%~\footnote{See also  Figs. 5.3 and 5.4 in Ref.~\cite{Brodsky:2014yha}, Figs. 2 and 3 in Ref.~\cite{Dosch:2015nwa},  and Fig. 1  in Ref.~\cite{Dosch:2015bca}.}. The agreement with experiment is generally better for the trajectories with total (cluster) spin $s=1$, such as the $\rho-\Delta$ superpartner trajectory  than for the trajectories with $s=0$, the $\pi-N$ trajectories. Features expected from spontaneous chiral symmetry  breaking are obtained, such as the masslessness of the pion in the massless quark limit. The structure of the superconformal algebra also implies that the pion has no supersymmetric partner. The meson-baryon supersymmetry survives, even if the heavy quark masses strongly break the conformal symmetry~\cite{Dosch:2015bca}.

The structure of the hadronic mass generation obtained from the supersymmetric Hamiltonian $G_S$,  Eq. \req{GS},  provides  a frame-independent decomposition of the quadratic masses for all four members of the supersymmetric multiplet. In the massless quark limit:
\beq 
M^2_H/ \lambda =\overbrace{\underbrace{(2n + L_H +1)}_{\it kinetic} +\underbrace{(2n + L_H +1)}_{\it potential }}^{\small  \begin{array}{c} \mbox{\it  contribution 
from 2-dim}\\  \mbox{\it light-front harmonic oscillator}\end{array}} \hspace{6mm}
+   \hspace{-6mm} \overbrace{2(L_H+ s)  +2 \chi}^{\small \begin{array}{c} \mbox{\it contribution 
from AdS and }\\  \mbox{\it superconformal algebra}\end{array}} .
\enq
Here $n$ is the radial excitation number and  $L_H$ the LF angular momentum of the hadron wave function; $s$ is the total spin of the meson and  the cluster respectively,   $\chi = -1$ for the meson and for  the negative-chirality component of the baryon (the upper components in the susy-doublet) and $\chi=+1$ for the positive-chirality component  of baryon and for the tetraquark (the lower components).  
The contributions  to the hadron masses squared  from the light-front  potential $\lambda^2 \zeta^2$ and the light-front kinetic energy 
in the LF Hamiltonian are identical because of the virial theorem.

We emphasize that the supersymmetric features of hadron physics derived here from superconformal quantum mechanics refers to the symmetry properties of  the bound-state wave functions of hadrons and not to quantum fields; there is therefore no need to introduce new supersymmetric  fields or particles such as squarks or gluinos.

We have argued that  tetraquarks -- which are degenerate with the baryons with the same (leading) orbital angular momentum-- are required to complete the supermultiplets predicted by the superconformal algebra. The tetraquarks are the bound states of the same confined color-triplet diquarks and anti-diquarks which account for baryon spectroscopy.

The light-front cluster decomposition~\cite{Brodsky:1983vf,Brodsky:1985gs} for a bound state of $N$ constituents  --as an ``active" constituent interacting with the remaining cluster of $N-1$ constituents-- also has implications for the holographic description of form factors.  As a result, the form factor is written as the product of a two-body form factor multiplied by the form factor of the $N-1$ cluster evaluated at its characteristic scale. The form factor of the $N-1$ cluster is then expressed recursively in terms of the form factor of the $N-2$ cluster, and so forth, until the overall form factor is expressed as the $N-1$ product of two-body form factors evaluated at different characteristic scales. This cluster decomposition is in complete agreement with the QCD twist assignment which leads to counting-rule scaling laws~\cite{Brodsky:1973kr, Polchinski:2001tt}. This solves a previous problem with the twist assignment  for the nucleon~\cite{Brodsky:2014yha}: The ground state solution to the Hamiltonian~\req{Gsd} for the nucleon corresponds to twist 2: the nucleon is effectively described by the wave function of a quark-diquark cluster.  At short distances, however, all of the constituents in the proton are resolved, and therefore the falloff of form factors at high $Q^2$ is governed by the total number of constituents; 
{\it i.e.}, it is twist 3~\footnote{A brief discussion of the LF cluster decomposition of form factors was given in Ref.~\cite{deTeramond:2016pov} and will be discussed in more detail elsewhere.}. 
Also the twist assignment for the $\Delta$ (and total quark spin-$\frac{3}{2}$ nucleons) deviates from the assignment introduced in our previous papers~\cite{deTeramond:2013it, Brodsky:2014yha, deTeramond:2014asa, Dosch:2015nwa};  the approach chosen here, dictated by supersymmetry, does not require the introduction of half-integer twist.

The emerging confinement mass scale $\sqrt \la$ serves as the fundamental mass scale of QCD; it is directly related to physical observables such as  hadron masses and radii;  in addition, as discussed in Ref.~\cite{Deur:2014qfa},  it can be related to the scheme-dependent perturbative QCD scales, such as the QCD renormalization parameter $\La_s$.

Qualitatively, the observed spectra of both light-quark baryons and mesons show approximately equally spaced parent and daughter trajectories with a common Regge slope~\cite{Klempt:2009pi}. This remarkable structure, especially the equal slopes of the meson and baryon trajectories, suggests the existence of a deeper underlying symmetry.   In the AdS/CFT correspondence~\cite{Maldacena:1997re} the dual quantum field theory is, in fact, a superconformal gauge theory.  Guided by these very general considerations, we 
have used a simple representation of superconformal algebra to construct semiclassical supersymmetric bound-state equations which are holographically mapped to relativistic  Hamiltonian bound-state equations  in the light front (LF)~\cite{Brodsky:2013ar, deTeramond:2014asa, Dosch:2015nwa}. These wave equations satisfactorily  reproduce  the successful empirical results previously obtained from light-front holographic QCD (see {\it e.g.}~\cite{Brodsky:2014yha}),  with the crucial advantage that additional constant terms in the confinement potential, which are essential for describing the observed phenomenology, are determined from the onset.

The supersymmetric approach  to hadronic physics also leads to unexpected connections  across the heavy-light hadron spectra,  a  sector where one cannot start from a superconformal  algebra because of the strong explicit breaking of conformal symmetry by heavy quark masses~\cite{Dosch:2015bca}.

In our framework, the emerging dynamical supersymmetry between mesons and baryons is not a consequence of supersymmetric QCD at the level of fundamental fields, but the  supersymmetry between the LF bound-state wave functions of mesons and baryons. This symmetry is consistent with an essential feature of color $SU(N_C)$:  a cluster of $N_C-1$ constituents  can be in the same color representation as the anti-constituent; for $SU(3)$ this means $\bf \bar 3 \in  \bf 3 \times \bf 3$ and $\bf  3 \in  \bf \bar3 \times \bf \bar3$~\footnote{This was the basis of earlier attempts~\cite{Miyazawa:1966mfa, Catto:1984wi, Lichtenberg:1999sc} to combine mesons and baryons in supermultiplets.}.

In   AdS$_5$  the positive and negative-chirality projections of the baryon wave functions, the upper and lower spinor components in the chiral representation of Dirac matrices, satisfy uncoupled second-order differential equations with degenerate eigenvalues.  These component wave functions form, together with the boson wave functions,  the supersymmetric multiplets.

The  semiclassical LF effective theory based on superconformal quantum mechanics also captures   other  essential features of hadron physics that one expects from confined quarks in QCD.  For example,  a massless  pseudoscalar $q \bar q$ bound state --the  pion--  appears in the limit of zero-quark masses,  and a mass scale emerges from a nominal conformal theory.  Moreover, the eigenvalues of the light-front Hamiltonian predict the same slope for Regge trajectories in both $n$,  the radial excitations,  and $L$, the orbital excitations, as approximately observed.  This nontrivial aspect of hadron physics~\cite{Glozman:2007at, Shifman:2007xn}  -- the observed equal slopes of the radial and angular Regge trajectories -- is  also a property of the Veneziano dual amplitude~\cite{Veneziano:1968yb}.

We  have derived a  semiclassical  light-front relativistic Hamiltonian based on superconformal algebra and its holographic embedding, which includes a spin-spin interaction between the hadronic constituents.  This approach leads to a remarkably simple, universal  and consistent description of the  light-hadron spectroscopy and their light-front wavefunctions.   We also predict the existence of tetraquarks which are degenerate with baryons with the same angular momentum. The tetraquarks are bound states of the same confined color-triplet diquark and anti-diquark clusters which account for baryon spectroscopy; they are required to complete the supermultiplet structure  predicted by the superconformal algebra.

\section*{Acknowledgments}

Presented at the Conference  ``New Physics at the Large Hadron Collider", 
29 February to 4 March 2016, at the 
Nanyang Executive Centre, Nanyang Technological University, Singapore.
We thank Alexandre Deur, Stanislaw D.  Glazek, Kelly Chiu, Xing-Gang Wu, and Matin Mojaza for valuable discussions.
This research was supported by the Department of Energy,  contract DE--AC02--76SF00515.  
SLAC-PUB-16545.

\end{document}